\documentclass[aps,nofootinbib,notitlepage,superscriptaddress,twocolumn,10pt,prd]{revtex4-1}

\usepackage{graphicx}
\usepackage{amsmath,amssymb}
\usepackage{hyperref}
\usepackage{braket}
\usepackage{physics}
\usepackage{subfigure}
\usepackage{float}
\usepackage[dvipsnames]{xcolor}
\usepackage{mathrsfs}
\usepackage{comment}
\usepackage{multirow}
\usepackage{tensor}
\usepackage{dsfont}
\usepackage[capitalise]{cleveref}
\usepackage{ulem}

\begin{document}

\title{
What does a cosmological experiment really measure? \\
Covariant posterior decomposition with normalizing flows
}

\author{Tara Dacunha} 
\author{Marco Raveri} 
\author{Minsu Park} 
\author{Cyrille Doux} 
\author{Bhuvnesh Jain} 
\affiliation{Center for Particle Cosmology, Department of Physics and Astronomy, University of Pennsylvania, Philadelphia, PA 19104, USA}

\begin{abstract}
We present methods to rigorously extract parameter combinations that are constrained by data from posterior distributions. The standard approach uses linear methods that apply to Gaussian distributions. We show the limitations of the linear methods for current surveys, and develop non-linear methods that can be used with non-Gaussian distributions, and are independent of the parameter basis. These are made possible by the use of machine-learning models, normalizing flows, to learn posterior distributions from their samples.
These models allow us to obtain the local covariance of the posterior at all positions in parameter space and use its inverse, the Fisher matrix, as a local metric over parameter space. The posterior distribution can then be non-linearly decomposed into the leading constrained parameter combinations via parallel transport in the metric space.  
We test our methods on two non-Gaussian, benchmark examples, and then apply them to the parameter posteriors of the Dark Energy Survey and Planck CMB lensing.
We illustrate how our method automatically learns the survey-specific, best constrained effective amplitude parameter $S_8$ for cosmic shear alone, cosmic shear and galaxy clustering, and CMB lensing. We also identify constrained parameter combinations in the full parameter space, and as an application we estimate the Hubble constant, $H_0$, from large-structure data alone.
\end{abstract}

\maketitle

\section{Introduction} \label{Sec:Introduction}

As the precision of cosmological surveys increases, so does the complexity of their analysis.
With the reduction of the statistical noise floor, new effects contaminating the cosmological signal are discovered and modeled to obtain robust constraints on cosmological parameters.
As a result, modern and future analyses need to properly deal with large number of parameters and their corresponding high dimensional parameter spaces.
The dimensionality of parameter spaces of present cosmological analyses already exceeds our visualization capabilities 
and make conceptually straightforward problems, like understanding the combinations of parameters that data is constraining, difficult in practice.

Prominent among cosmological surveys are temperature and polarization maps of the cosmic microwave background (CMB)~\cite{Planck:2018nkj, SPT:2019fqo, ACT:2020gnv} and galaxy surveys carried out with optical and near-infrared telescopes~\cite{Heymans:2020gsg, Hamana:2019etx, DES:2021wwk}. 

CMB surveys are used for cosmological analysis of the early universe, at redshift $z\approx 1100$, while galaxy surveys map the expansion history and large scale structure at late times, typically $0<z\lesssim1$. 
In recent years, two-point statistics such as the power spectrum of fluctuations have been used for parameter analysis. While CMB measurements probe a Gaussian field, galaxy surveys map the non-linear structure of the late time universe, which requires statistics beyond two-point correlations~\cite{DES:2017hhj,DES:2021epj,DES:2021lsy,Harnois-Deraps:2020pvj,Appleby:2021xoz} to fully characterize the density field (probed using tracers such as the galaxy number density or the weak lensing shear). This is because the density fluctuations have evolved via non-linear gravitational clustering, which produces a non-Gaussian field, especially on small scales. 

Related to the non-Gaussianity of the density field is that the cosmological parameter combinations best probed by galaxy surveys depend on the details of the analysis, in particular: 1. The length scales included in the analysis, and how deeply into the non-linear regime they extend, 2. The redshift coverage, and 3. The choice of statistic used to extract information from the survey data. Specifically, if 3-point correlations~\cite{Ivanov:2021kcd} or other statistics that go beyond 2-point correlations are used, especially in combination, then the parameter combinations probed can vary significantly~\cite{Barreira:2021ueb,Gualdi:2021yvq,Jain:1996st, Bernardeau:1996un, Takada:2003ef}. As an example, 2-point correlations of the weak lensing shear field are widely characterized by two leading parameters, $S_8\equiv \sigma_8 \Omega_m^{0.5}$ and $\Omega_m$. The $S_8$ parameter is used because it is the best constrained direction in the $\sigma_8-\Omega_m$ parameter space, and constraints in the $S_8-\Omega_m$ space are close to orthogonal. However, higher order statistics are sensitive to different weightings of $\sigma_8$, $\Omega_m$ and possibly other parameters~\cite{Pyne:2020ijd,ChengChengSiHao:2021hja,Halder:2021itp,DES:2019ujq}. This poses a challenge in how best to interpret the extracted cosmological information, especially when comparing constraints from different surveys.

Statistical tensions among different datasets can signal new physics and/or problems in the analysis of one or more of the datasets. Recent work has explored different ways of computing the consistency or tension among datasets in the full parameter space of interest~\cite{Raveri:2018wln, Raveri:2019gdp, Handley:2019wlz, Park:2019tyw, DES:2020hen}.
For practical purposes, it is often convenient to consider a restricted part of the parameter space, ideally in two or even one dimension for visualization purposes. Such a step requires resolving the issue discussed above: different statistics and surveys are sensitive to different parameter combinations. 

In this paper, we develop the methods to understand what is being constrained by data from cosmological surveys.
In doing so, we follow the guiding principle of general covariance.
The physical laws underlying our data do not depend on how we write it down, hence results that we derive from posterior distributions should not depend on the parameter basis that we choose.
We show that the commonly used Principal Component Analysis (PCA)~\cite{doi:10.1080/14786440109462720} does not respect this principle, hence we modify it developing Covariant Principal Component Analysis (CPCA).
We derive the properties of CPCA in cases of Gaussian distributions where a decomposition of parameter space correspond to a set of linear projections.
We then use machine-learning models, namely normalizing flows, to extend CPCA to deal with non-Gaussian distributions.
After applying these techniques to benchmark examples, we show an end-to-end application to cosmology, considering data from the Dark Energy Survey~\cite{DES:2017myr} and Planck CMB lensing reconstruction~\cite{Planck:2018lbu}.

This paper is organized as follows:
in \cref{Sec:LinearMethods}, we discuss linear methods that apply to Gaussian distributions, reviewing PCA in \cref{Sec:PCA} and developing CPCA in \cref{Sec:KL};
in \cref{Sec:NonLinearMethods}, we derive non-linear methods that apply to non-Gaussian distributions;
in \cref{Sec:ToyExamples}, we show the application of both PCA and CPCA to benchmark non-Gaussian examples;
in \cref{Sec:RealExamples}, we discuss the application to cosmological surveys with a set of examples. 

\section{Linear methods} \label{Sec:LinearMethods}
We start by setting up some notation that we are going to use hereafter.
We indicate with $\theta$ the parameters of model $\mathcal{M}$, and $D$ the measured data at hand.
Then the posterior distribution is defined as the probability distribution of model parameters given the data and model, $\mathcal{P}(\theta) \equiv P(\theta|D, \mathcal{M})$, and given by Bayes theorem as:
\begin{align} \label{Eq:Posterior}
\mathcal{P}(\theta) = \frac{\mathcal{L}(\theta)\Pi(\theta)}{\mathcal{E}} \,,
\end{align}
where the prior distribution is defined as the probability of model parameters given the model regardless of the data $\Pi(\theta) \equiv P(\theta|\mathcal{M})$;
the likelihood is defined as the probability of the data given a model and a choice of parameters $\mathcal{L}(\theta) \equiv P(D|\theta, \mathcal{M})$;
the normalization of the posterior, the evidence, is the probability of the data given the model $\mathcal{E} \equiv P(D|\mathcal{M}) = \int \mathcal{L}(\theta) \Pi(\theta) \, d\theta$.
In the following we suppress explicit dependence on the model since we always operate at fixed model.

We indicate quantities that refer to the posterior distribution with the subscript $p$, the prior distribution with the subscript $\Pi$, and the likelihood with the subscript $d$. 
We denote a Gaussian distribution over parameters $\theta$, with mean $\hat{\theta}$ and covariance $\mathcal{C}$ as $\mathcal{N}_\theta(\hat{\theta}, \mathcal{C})$.

In this section, we assume that the posterior, likelihood, and prior are Gaussian distributions in parameter space with mean and covariance given by $\theta_p$, $\mathcal{C}_{p}$, $\theta_d$, $\mathcal{C}_{d}$, $\theta_\Pi$, $\mathcal{C}_{\Pi}$
respectively.
We indicate the inverse covariance, the Fisher matrix, with $\mathcal{F} \equiv \mathcal{C}^{-1}$.
While this is generally an approximation, for instance when the true prior is flat, a Gaussian distribution still captures two key aspects of a potentially informative prior, i.e. changing the center and scale of the posterior with respect to the likelihood.

Since we are working with Gaussian distributions we can write the (inverse of the) posterior covariance as:
\begin{align} \label{Eq:Cp_Def}
\mathcal{C}_p^{-1} = \mathcal{C}_d^{-1} + \mathcal{C}_\Pi^{-1},
\end{align}
and the posterior mean/maximum as:
\begin{align} \label{Eq:Thetap_Def}
\theta_p = \mathcal{C}_p ( \mathcal{C}_d^{-1}\theta_d + \mathcal{C}_\Pi^{-1} \theta_\Pi),
\end{align}
both in terms of the likelihood and prior means and covariances.

To facilitate the comprehension of this section, we follow along the text with a simple example of a Gaussian distribution in two correlated variables.
We also use, when needed, a Gaussian prior that is weakly informative.
This example is selected to match the Gaussian approximation of the real example discussed in \cref{Sec:RealExamples}.

\subsection{Posterior Principal Component Analysis} \label{Sec:PCA}
We start by discussing the commonly used procedure of Principal Component Analysis (PCA)~\cite{doi:10.1080/14786440109462720}.

This amounts to computing the eigenvalues and eigenvectors of the posterior inverse covariance matrix or Fisher matrix, thus solving:
\begin{align} \label{Eq:PCA}
\mathcal{F}_p Q = Q \Lambda
\end{align}
where $Q$ is an orthogonal matrix ($Q^{-1} = Q^T$) whose columns are the eigenvectors. 
Since $\mathcal{F}_p$ is real and symmetric then $\Lambda$ is a diagonal matrix with positive eigenvalues $\lambda_i$.
In this context, we could use both the posterior or likelihood Fisher matrices, and conclusions of this section would apply similarly.

We note the following immediate properties:
\begin{enumerate}
\item $\mathcal{F}_p = Q \Lambda Q^T$ so that the eigenvector basis completely decomposes the Fisher information matrix;
\item $\mathcal{C}_p = Q \Lambda^{-1} Q^T$ where $\Lambda^{-1}$ is diagonal with $1/\lambda_i$ on the diagonal. This means that the PC modes for the covariance and Fisher matrix are the same.
\item if we define $q \equiv Q^T \theta$, then ${\rm Cov}(q) = \Lambda^{-1}$, as a consequence of $Q^T \mathcal{C}_p Q = \Lambda^{-1}$ and the $q$ coordinates are statistically independent since $\Lambda$ is diagonal.
\end{enumerate}
It can be shown~\cite{ghojogh2019eigenvalue} that we can write PCA as an optimization problem:
\begin{align} \label{Eq:PCAOptimization}
(\mbox{PCA}) \ \ \ &\mbox{maximize} \enspace \theta^T \mathcal{C}_p^{-1} \theta \nonumber \\
&\mbox{subject to} \enspace \theta^T\theta = 1
\end{align}
whose solution are the eigenvectors of the Fisher matrix.
In this context, we note that the same optimization problem can be similarly written for the covariance instead of its inverse since they share the same eigenvectors, up to a redefinition of the eigenvalues.

\subsubsection{PCA and reparametrizations}

We could now be tempted to sort modes by their eigenvalues and consider the parameter combinations corresponding to the largest eigenvalues as the best constrained modes, with more information than the other modes.
There is one major roadblock to this procedure, which is that PCA is not covariant under a linear transformation of parameters.

By this, we mean that we consider a linear reparametrization:
\begin{align} \label{Eq:LinearReparametrization}
\tilde{\theta} = A \theta
\end{align}
where $A$ is an invertible base change matrix. 
This transformation clearly carries no physical meaning since the physics behind the data constraints we are considering has to be independent on how we write it down.
In general, $A$ rotates and rescales the values of the parameters.
This class of transformations also preserves Gaussianity of the distributions we are considering.
Under such a transformation, the PCA eigenvalues will generally change and the PC modes in the new basis will not be the transformed PCA directions in the original basis.

To see this consider that, under the transformation in \cref{Eq:LinearReparametrization}, the parameter Fisher matrix changes to:
\begin{align} \label{Eq:Cp_tilde}
\tilde{\mathcal{F}}_p = A^{-T} \mathcal{F}_p A^{-1} \,.
\end{align}
We can then rewrite the eigenvalue problem in \cref{Eq:PCA} as:
\begin{align}
A^{T} \tilde{\mathcal{F}}_p A\, Q = Q \Lambda
\end{align}
which can be simplified to the form:
\begin{align}
\tilde{\mathcal{F}}_p [A Q] = [A^{-T} Q] \Lambda,
\end{align}
that is, if $Q$ are eigenmodes of $\mathcal{F}_p$ then generally $AQ$ or $A^{-T}Q$ are not eigenmodes of $\tilde{\mathcal{F}}_p$, unless $A^{-T} = A$, i.e. $A$ is an orthogonal matrix and the base change transformation is a global isometry. 
Since the base change matrix $A$ is not necessarily orthogonal, PCA is not generally covariant under an invertible, linear reparametrization.

Note also that, while the PC modes $Q$ are orthonormal in their original parameter basis, they are not once transformed to the new parameter basis, so the decorrelation of parameters provided by PCA only holds in a specific parameter basis.

This means that the principal components of the covariance matrix are as arbitrary as the parameter basis that is used to compute them.
In particular, PCA applied to the parameter posterior covariance matrix would depend on the physical units of the parameters, as a units change is not a linear isometry.
To mitigate this problem PCA is usually performed on the posterior correlation matrix that would, at least, solve the problem with units choice although not solving the general problem of lack of linear covariance. In addition, the PC modes of the covariance and correlation matrix would not be the same, in a fixed parameter basis.

If we consider PCA as an optimization problem, as in \cref{Eq:PCAOptimization}, we clearly see why the method is not linearly covariant. While the optimization target is invariant under a linear transformation, the constraint is not.

\begin{figure}
\includegraphics[width=\columnwidth]{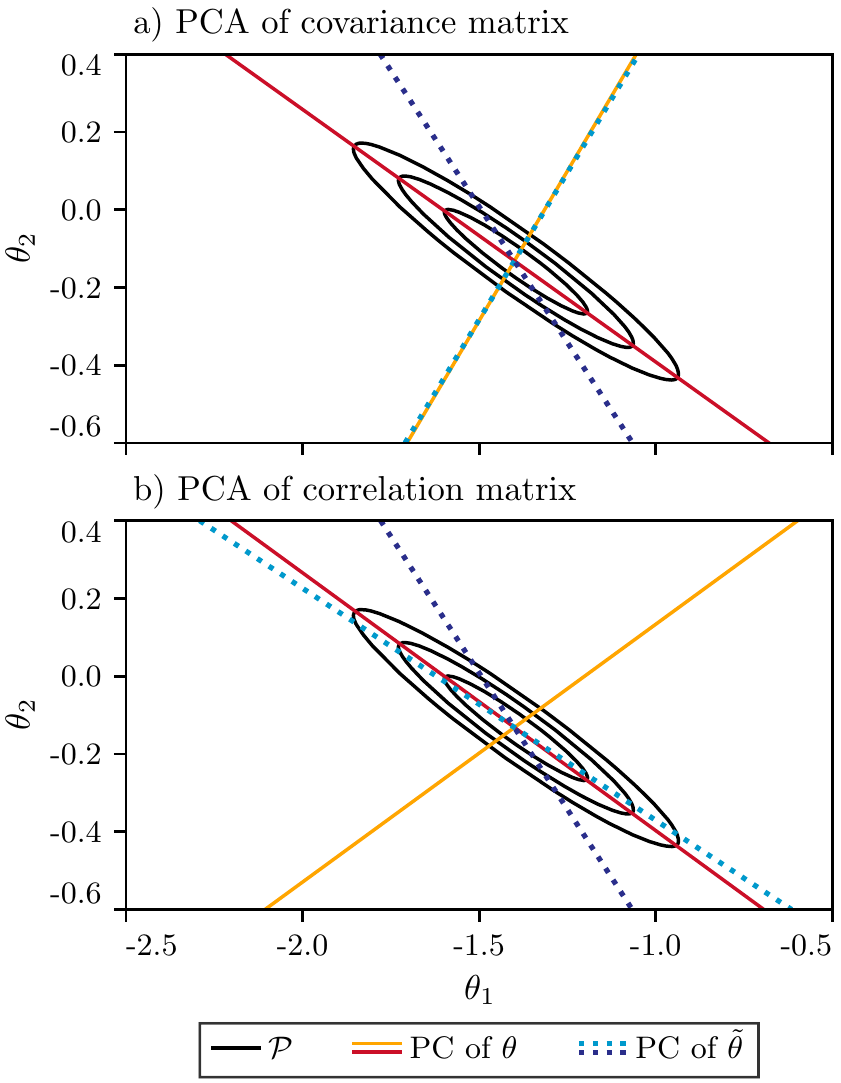}
\caption{ \label{fig:gaussian_example_PCA}
A toy example of how principal components (PCs) differ depending on the parameter basis chosen. The upper panel uses the covariance or Fisher matrix to determine the principal components while the lower panel uses the correlation matrix. Both panels display the difference between calculating the principal components in the original and linearly transformed basis as in \cref{Eq:transformation}.
}
\end{figure}

In \cref{fig:gaussian_example_PCA} we use a toy example to demonstrate the lack of covariance of PCA under a linear transformation of parameters. 
As in \cref{Eq:LinearReparametrization}, we transform parameters $\theta_1$ and $\theta_2$ with a linear transformation described by the matrix: 
\begin{align} \label{Eq:transformation}
A = 
\begin{bmatrix}
1 & -1\\
0 & 1
\end{bmatrix}
\end{align}
that does not transform the second parameter but changes the first parameter to be the difference between the two original parameters.
The Fisher matrix $\mathcal{F}$ transforms as in \cref{Eq:Cp_tilde} into $\tilde{\mathcal{F}}$. We calculate the principal components of $\mathcal{F}$ and  $\tilde{\mathcal{F}}$. The principal components of $\tilde{\mathcal{F}}$ are then transformed back to original parameter space. 
Both sets of principal components are shown in Panel a) of \cref{fig:gaussian_example_PCA} and are visibly distinct from each other. We apply the same process to the correlation matrix, resulting in principal components that are different from the ones computed from the covariance and that differ in the two parameter bases that we consider.

Each PC mode $i$ defines a direction in parameter space that we can write starting from its orthogonal plane:
\begin{align} \label{Eq:PlanePCA}
Q_i^T (\theta -\theta_p) = 0 \pm \lambda_i^{-1/2},
\end{align}
where $Q_i$ is the $i$-th column of the matrix $Q$.
This inner product would have as many terms as parameters involved and it is generally of little practical use.
We can make these expressions more compact by neglecting some parameters.
We start by noticing that we can write \cref{Eq:PCA} as:
\begin{align}
\qty(\sqrt{\mathcal{F}_p} Q)^T \qty(\sqrt{\mathcal{F}_p} Q) = \Lambda,
\end{align}
since $\mathcal{F}_p$ is symmetric and positive definite, and hence has a symmetric matrix square root.
We can then define, for a given mode $i$:
\begin{align}
T_{ij} = \frac{\qty(\sqrt{\mathcal{F}_p} Q)_{ij}^2}{\lambda_i}
\end{align}
which quantifies the fractional contribution of parameter $j$ to the variance of the PC mode $i$.
By virtue of \cref{Eq:PCA}, this has to sum to one, $\sum_j T_{ij} = 1$.

From the implicit expression for a given PC mode, \cref{Eq:PlanePCA}, we can neglect parameters if they are not really contributing to the mode variance, up to a set relative threshold.
This compression does break all properties of PCA but it proves useful and insightful at times.

In \cref{Sec:PCA.SNR} we further discuss properties of PCA related to signal-to-noise ratio in parameter space.

\subsection{Data and parameters: Karhunen-Loeve decomposition}
We could then try to build linear decomposition methods that are a bridge between data and parameter space, similar to what is discussed in~\cite{Tegmark:1996bz}.

We start by assuming that the data $x$ is Gaussian distributed with covariance $\Sigma$ around model predictions $m(\theta)$.
We expand to linear order model predictions around a point  that we set to the prior center for convenience:
\begin{align}
m(\theta) = & m(\theta_\Pi) + M(\theta - \theta_\Pi) +\cdots
\end{align}
where $M=\partial m / \partial \theta |_{\theta_\Pi}$ is the model prediction Jacobian.
We can then define:
\begin{align}
\tilde{M} = \mathcal{C}_d M^T\Sigma^{-1}
\end{align}
where 
\begin{align} \label{Eq:GLMParameterCovariance}
\mathcal{C}_d = (M^T\Sigma^{-1}M)^{-1},
\end{align}
and then we have:
\begin{align}
\theta_d -\theta_\Pi = \tilde{M}(x-m_\Pi)
\end{align}
That is just saying that, in the Gaussian linear model, the best-fit parameters are a linear combinations of the data.

We are now in a position to define hybrid decompositions.
Since we are interested in parameter combinations, we can define the following optimization problem:
\begin{align} \label{Eq:KLdataOptimization}
(\mbox{KL}) \ \ \ &\mbox{maximize} \enspace \theta^T \mathcal{C}^{-1}_d \theta \nonumber \\
&\mbox{subject to} \enspace x^Tx = \theta^T M^T M \theta = 1.
\end{align}
This boils down to the calculation of the generalized eigenvalues of $\mathcal{F}_d=\mathcal{C}^{-1}_d$ and $M^T M$.
This decomposition is also called the Karhunen-Loeve (KL) decomposition in~\cite{Tegmark:1996bz, Raveri:2018wln}.
In writing the second line of \cref{Eq:KLdataOptimization}, we have used the fact that $M\tilde{M}$ defines a projector on the model tangent space, as discussed in~\cite{Raveri:2018wln}, which selects the only ways in which a data mode can change parameter inference.

The decomposition in \cref{Eq:KLdataOptimization} is linearly covariant in parameter space.
However, it is not covariant under linear transformations of the data and in general it would depend on the data units.
This is a problem once we consider different data points with different units, as in the case of two different data sets.
These will in general have different scales or different units that cannot be easily compared to one another.

In general the data modes that contribute a constraint are affine covariant in both parameter and data space. However, the way in which their constraint is distributed over different parameters in the previous decomposition is not.
This is the roadblock to generalizing the discussion in~\cite{Tegmark:1996bz} to more than one parameter at a time.

If we wanted to make the above problem fully affine covariant in both parameter and data space, then we would need to weight the data constraint by the inverse covariance of the data:
\begin{align} \label{Eq:KLdataOptimization2}
&\mbox{maximize} \enspace \theta^T \mathcal{C}^{-1}_d \theta \\
&\mbox{subject to} \enspace x^T\Sigma^{-1}x = \theta^T M^T \Sigma^{-1} M \theta = \theta^T \mathcal{C}^{-1}_d \theta = 1 \nonumber
\end{align}
which, however, by \cref{Eq:GLMParameterCovariance}, makes the problem trivial.
This happens because, within the linear model, the parameter likelihood covariance is just the model projection of the data covariance.

\subsection{Covariant Principal Component Analysis} \label{Sec:KL}
To make progress, we note that in general for Bayesian inference the parameter posterior covariance and the model projection of the data covariance are not the same. 
If we modify the optimization problem in \cref{Eq:KLdataOptimization2} to:
\begin{align} \label{Eq:KLdataOptimization3}
(\mbox{CPCA}) \ \ \ &\mbox{maximize} \enspace \theta^T \mathcal{C}^{-1}_p \theta \nonumber \\
&\mbox{subject to} \enspace x^T\Sigma^{-1}x = \theta^T \mathcal{C}^{-1}_d \theta = 1,
\end{align}
then this does have non-trivial solutions and is fully linearly covariant in both data and parameter space.

This optimization problem corresponds to the KL decomposition of the posterior and data Fisher matrices~\cite{ghojogh2019eigenvalue} and it can be shown that it is equivalent, up to a redefinition of the generalized eigenvalues, to solving the generalized problem:
\begin{align} \label{Eq:KL}
\mathcal{F}_{p} \Psi = \mathcal{F}_\Pi \Psi \Lambda \,,
\end{align}
where $\Psi$ are KL modes and $\Lambda$ is a diagonal matrix.
This form of the KL decomposition is more useful in practice, with respect to \cref{Eq:KLdataOptimization3}, since both matrices involved can be easily obtained from parameter samples. From this point on, we refer to this KL decomposition method as Covariant Principal Component Analysis (CPCA). 

Some useful properties of the CPCA hold:
\begin{align} \label{Eq:DecorrelateKLFisher}
\Psi^T \mathcal{F}_\Pi \Psi =& I \nonumber \\
\Psi^T \mathcal{F}_p \Psi =& \Lambda \nonumber \\
\Psi^T \mathcal{F}_d \Psi =& \Lambda - I \,.
\end{align}
These show that the CPC modes diagonalize and decorrelate jointly the prior, data and posterior distributions, all the Fisher matrices that we have available in the linear setup.
The first equation shows that the prior is setting the units for the decomposition problem, the second clarifies the meaning of the CPC eigenvalues, as gauging the information improvement of the posterior with respect to the prior.
The third equation follows from the fact that, in the linear model, the data Fisher matrix can be obtained from the other two, hence a transformation that diagonalizes two of them also diagonalizes the third one.
In this sense, the CPCA of the prior and posterior is unique, as it is the transformation that diagonalizes all covariances we have available in the linear model.
We may compute the generalized eigenvalue decomposition of all pairs of covariance matrices, and we would always find the same directions, modulo a redefinition of $\Lambda$.

Unlike PCA, this CPCA is covariant under linear parameter transformations. 
Consider the same linear reparametrization as in \cref{Eq:LinearReparametrization}.
The prior Fisher matrix transforms in the same way as the posterior Fisher matrix so that \cref{Eq:KL} becomes:
\begin{align}
A^T \tilde{\mathcal{F}}_{p} A \Psi = A^{T} \tilde{\mathcal{F}}_\Pi A \Psi \Lambda \,,
\end{align}
which is simply:
\begin{align}
\tilde{\mathcal{F}}_{p} [A \Psi] = \tilde{\mathcal{F}}_\Pi [A \Psi] \Lambda \,.
\end{align}
This means that CPC modes in a parameter basis are also CPC modes in another basis under the coordinate transformation $\tilde{\Psi} = A \Psi$.
This shows that the CPC modes of the posterior, with respect to the prior are covariant.
Most interestingly the CPC eigenvalues, $\Lambda$, that quantify the improvement of the posterior over the priors are parameter invariants.
Due to the transformation laws of both the Fisher matrix and the KL modes, \cref{Eq:DecorrelateKLFisher}, also shows that CPCA decorrelates parameters in every parameter basis.

\begin{figure}% [!ht]
\includegraphics[width=\columnwidth]{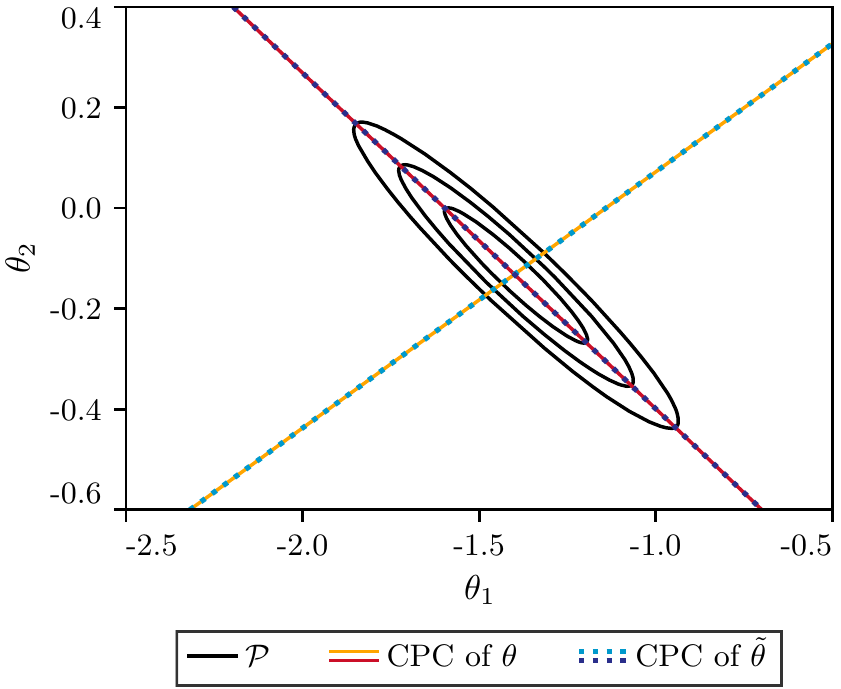}
\caption{ \label{fig:gaussian_example_KL}
A toy example of how the Covariant Principal Components (CPC's) of the Fisher matrix are invariant under linear transformations of parameters, as in \cref{Eq:transformation}.
This figure can be compared to the analogous  \cref{fig:gaussian_example_PCA} for PCA.
}
\end{figure}

In \cref{fig:gaussian_example_KL} we use the same toy example of \cref{fig:gaussian_example_PCA} to 
demonstrate the covariance of CPCA under a linear transformation of parameters. 
As we can see, regardless of the parameter basis that is used, both sets of CPC's are indistinguishable from each other. 

We report here some relations that are sometimes useful in practice:
\begin{align} \label{Eq:CPCA.decomposition.Fisher}
\mathcal{F}_\Pi =& \Psi^{-T} \Psi^{-1} \,, \nonumber \\
\mathcal{F}_p =& \Psi^{-T} \Lambda  \Psi^{-1} \,, \nonumber \\
\mathcal{F}_d =& \Psi^{-T} (\Lambda  - I) \Psi^{-1} \,,
\end{align}
that allow to write all relevant Fisher matrices in terms of CPC modes.
From these we can see that CPC modes decompose the covariance matrix as well:
\begin{align} \label{Eq:CPCA.decomposition.Covariance}
\mathcal{C}_\Pi =& \Psi  \Psi^T \,, \nonumber \\
\mathcal{C}_p =& \Psi \Lambda^{-1} \Psi^T \,, \nonumber \\
\mathcal{C}_d =& \Psi (\Lambda  - I)^{-1} \Psi^T \,.
\end{align}
These also clarify the relation between the CPC modes of the Fisher matrix and the posterior:
if $\Psi$ are CPC modes of the posterior/prior Fisher matrices then $\Psi^{-T}$ are the CPC modes of the same covariance matrices.
\Cref{Eq:CPCA.decomposition.Fisher,Eq:CPCA.decomposition.Covariance} also allow to obtain, in a numerically stable way, the Fisher and covariance matrices of the likelihood, $\mathcal{F}_d$ and $\mathcal{C}_d$, that we would not be able to obtain from samples otherwise.
We can also see from 
\cref{Eq:CPCA.decomposition.Covariance} that $\sqrt{\lambda-1}$ quantifies the variance improvement of the data over the prior.

As a consequence, for posterior parameters, when we project them on CPC modes, i.e. $K_p=\Psi^{-1} \theta_p$, then 
$K_p$ are uncorrelated with covariance $\Lambda^{-1}$.
This is analogous to the projection to PC in the previous section, with the difference that the covariance of the projected CPC modes does not depend on the parameter basis.

As for PCA in the previous section, a given CPC mode defines a hyper-plane that is perpendicular to it:
\begin{align} \label{Eq:PlaneKL}
\Psi_i^{-1} (\theta -\theta_p) = 0 \pm \lambda_i^{-1/2} \,,
\end{align}
and we can define the relative contribution of parameter $j$ to the variance of the mode $i$ by:
\begin{align}
T_{ij} = \frac{(\sqrt{\mathcal{F}_p} \Psi)_{ij}^2}{\lambda_i} \,,
\end{align}
and remove parameters from the implicit form of a CPC mode in \cref{Eq:PlaneKL} if they are not contributing to the variance of the mode to a given threshold.

In \cref{Sec:KL.SNR} we discuss properties of CPCA related to the decomposition of signal-to-noise ratio and to the Kullback–Leibler divergence.

\section{Non-linear methods} \label{Sec:NonLinearMethods}
In this section, we discuss non-linear methods that apply to non-Gaussian distributions, with possibly non-Gaussian priors.

To facilitate the comprehension of this section we follow along the text with a simple example of a log-normal distribution in two variables, with flat uninformative priors.
This example is selected to match the log-normal approximation of the real example discussed in \cref{Sec:RealExamples}.

\subsection{Normalizing flows} \label{Sec:NormalizingFlows}
The starting point for the non-linear methods we discuss in this section consists in having a model for the posterior distribution.
This is provided by normalizing flows, that are generative models with tractable distributions based on invertible and differentiable transformations. 
Given samples drawn from an unknown distribution, they can model that distribution by learning a mapping from the sample space to an abstract space where the transformed samples have a simple analytic distribution, such as a unit Gaussian distribution, \cite{2017arXiv170507057P,2018arXiv180703039K,2018arXiv181001367G,2019arXiv190809257K,2020arXiv200301941M}.
They have recently found applications in astrophysics for modeling \cite{2020arXiv200600615R,2021arXiv211002983H}, cosmological parameter inferences \cite{2020PhRvD.102j3507D,2020MNRAS.496..328M,2020MNRAS.tmp.3445J} and the quantification of tensions between cosmological constraints inferred from different measurements \citep{Raveri:2021wfz}.

For a point in parameter space $\theta$ and a mapping $f$ from a unit Gaussian distribution with density $\varphi$, the probability density in parameter space is approximated by:
\begin{align}
    p(\theta) \approx \varphi\qty(f^{-1}(\theta)) \det J_{f^{-1}},
\end{align}
where $\det J_{f^{-1}}$ is the determinant of the Jacobian of $f^{-1}$. The mapping may be parametrized by neural networks that are trained to maximize the similarity between the left and right sides of the above. In other words, the loss function is given by the Kullback-Leibler divergence between $p(\theta)$  and the learned distribution, which is equivalent to $-\log \varphi\qty(f_\alpha^{-1}(\theta))$ up to a constant term, where $\alpha$ represents the weights and biases of the neural networks. The density $p(\theta)$ can then be computed using automatic differentiation.

Following~\cite{Raveri:2021wfz}, we use Masked Autoregressive Flows \citep[MAF,][]{2017arXiv170507057P} to learn posterior distributions $\mathcal{P}(\theta)$ from samples. MAFs are composed of a stack of neural networks called Masked Autoencorders for Density Estimation \cite[MADE,][]{2015arXiv150203509G}, that each implement a simple autoregressive transformation. For an input vector $x$ with $n$ components $x_i$, ${i=1 \dots n}$, the output vector $y$ of an autoregressive transformation has components given by $y_1=x_1$ and ${y_i=\sigma(x_{1\dots i-1})x_i+\mu(x_{1\dots i-1})}$ for ${i=2 \dots n}$, where $\mu$ and $\sigma$ represent unconstrained neural networks that receive the masked input ${x_{1\dots i-1}=(x_1,\dots,x_{i-1},0,\dots,0)}$. Stacking such simple transformations with random permutations of components in between allows MAFs to have higher expressive power than each MADE individually, which permits to model many different distributions.

For the problem of modeling posterior distributions that derive from a known prior distribution, we further decompose the mapping of the normalizing flow as follows:
\begin{enumerate}
    \item We first use a transformation that maps the prior distribution to a unit Gaussian distribution. If the prior distribution $\pi$ is a known, analytical distribution such as a combination of uniform and multivariate Gaussian priors, we compute the analytic transformation, $f_\pi$, from a unit Gaussian distribution to the prior space using (inverse) cumulative distribution functions. Otherwise, we may first learn the prior distribution itself using prior samples. In this case, we follow the same implementation as described here, assuming a uniform distribution within the minimum and maximum values allowed for the parameters.
    \item Next, we compute the empirical mean and covariance of the transformed samples, $f^{-1}_\pi(\theta)$, which we use to define the transformation $f_\odot$ that maps a unit Gaussian distribution to a multivariate Gaussian distribution with this mean and covariance.
    \item Finally, we train a MAF to learn the mapping $f_{\rm MAF}$ from a unit Gaussian to the distribution of the (twice) transformed samples $f^{-1}_\odot\qty(f^{-1}_\pi(\theta))$.
\end{enumerate}
The full transformation from a unit Gaussian to the posterior space is therefore given by the composition of these three transformations, ${f = f_\pi \circ f_\odot \circ f_{\rm MAF}}$, such that the transformed samples $f^{-1}_{\rm MAF}\qty(f^{-1}_{\odot}\qty(f^{-1}_{\pi}(\theta)))$ are distributed as a unit Gaussian. The first two transformations are fixed by the knowledge of the prior and by the posterior sample at hand, while the third one requires optimization.

The motivation for this decomposition is the following. Since we know the prior, applying the transformation that gaussianizes the prior allows to correctly account for (potentially) informative boundaries of uniform priors and to restrict the normalizing flow to the corresponding part of parameter space. In our tests, this has proven to be very efficient. The second transformation allows us to pre-gaussianize the transformed posterior samples, thus simplifying the mapping that remains to be learned by the MAF.

As a strategy to improve the performances of the flow model and mitigate cases where we would have a bad initialization of weights, we train several flow models starting from different random weight initializations.
We then pick the model that has the best loss on the validation set of samples.

\subsection{Bayesian information geometry with normalizing flows} \label{Sec:IGwithNormalizingFlows}
We now seek to define a geometric structure on parameter space.
To do so, we start from the usual definition of the Fisher information metric:
\begin{align} \label{Eq:Fisher.definition}
\mathcal{F}_{ij}(\theta) &\equiv - \int \dd{x} \ \mathcal{L}(x|\theta) \partial_{ij}\log \mathcal{L}(x|\theta),
\end{align}
where the integral is performed over data realizations, $x$.

This definition, however, does not allow us to build such a structure for prior or posterior distributions.
Indeed, once samples from the posterior are obtained, any connection with the distribution of data is lost.
Moreover, prior distributions have no connection to underlying data at all.
The definition in \cref{Eq:Fisher.definition} is also impractical in realistic scenarios.

We start with a probability density, $p(\theta)$, with no explicit reference to data and assume that we can Gaussianize it using
normalizing flows to learn the mapping between coordinates, $\tilde{\theta} \equiv \phi(\theta) = f^{-1}(\theta)$, such that $\tilde{\theta}$ has a unit Gaussian distribution. 
As discussed in the previous section, we find MAF normalizing flows to have enough expressive power to always allow for this to happen in realistic cases.
We can now generate a family of distributions that is needed to compute the local Fisher matrix, with \cref{Eq:Fisher.definition}, by translations in abstract space.
To do so, we assume that we can generate new data, $\tilde{\theta}^\prime$, as realizations of parameters, so that their distribution is given by $\mathcal{N}_{\tilde{\theta}^\prime}(m(\tilde{\theta}), \mathds{1})$, and we assume that ${m(\tilde{\theta}) = \tilde{\theta}}$.

Then the Fisher matrix is given by the identity matrix:
\begin{align} \label{Eq:AbstractSpaceFlat}
    \tilde{\mathcal{F}}_{ij}(\tilde{\theta}) &= - \int \dd{\tilde{\theta}'} p(\tilde{\theta}'|\tilde{\theta}) \left(\frac{\partial^2}{\partial \tilde{\theta}^i\partial \tilde{\theta}^j} \log p(\tilde{\theta}'|\tilde{\theta}) \right)\\
    &= - \int \dd{\tilde{\theta}'} p(\tilde{\theta}'|\tilde{\theta})  (-\delta_{ij}) = \delta_{ij},
\end{align}
where $\tilde{\theta}^\prime$ is the data being integrated over, and $\tilde{\theta}$ is the parameter. 
As a consequence of \cref{Eq:AbstractSpaceFlat}, the abstract parameter space is globally flat.
Once we have the Fisher matrix in abstract space, the Fisher matrix for the posterior in original space is given by its tensor transformation rule:
\begin{align}
\mathcal{F}_{\mu\nu} &= \frac{\partial \phi^{a}} {\partial {\theta}^\mu} \delta_{ab}  \frac{\partial \phi^{b}}{\partial {\theta}^\nu} \,,
\end{align}
where, hereafter, we assume the Einstein summation convention that repeated indices are summed over.

Since we have a global mapping, $\phi$, from our original parameter space to a flat space, the original parameter space is flat as well, and normalizing flows introduce a reparametrization which does not change the geometry itself.
This is a consequence of demanding that data realizations are parameter realizations and is the only way in which we can define a Fisher matrix for prior distributions where we do not have data.
As a consequence, both the Riemann and Ricci curvature tensors vanish.

Note that the Jacobian $\partial \phi^{a}/\partial {\theta}^\mu$ learned by  normalizing flows is not unique, because  normalizing flows simply learn the mapping required for gaussianization without boundary conditions. However, this redundancy in the Jacobian is irrelevant as the metric $\mathcal{F}_{\mu\nu}$ is unique. 
From the fact that $\mathcal{F}$ is symmetric and $\partial \phi^{a}/\partial {\theta}^\mu$ is not, we can understand, heuristically, that the redundancy of the non-uniqueness is cancelled away in multiplying two Jacobians to get the symmetric metric. 

We can connect this definition to the standard definition of the Fisher matrix that is generally used in information geometry~\cite{10.5555/3019383}.
In this context, all prior distributions are flat spaces, since they have no underlying data beyond parameters.
On the other hand, for likelihoods, our treatment can be thought of as finding Riemann normal coordinates around the observed data realization.
In this sense, we note that the standard formulation of information geometry conflicts with the principle of Bayesian updating.
In Bayesian statistics, there exists a symmetry between the prior and the posterior, i.e. we can use the result of a previous experiment as priors for the next experiment. 
This is not true in information geometry, since the geometry of the two distributions would be different, flat in one case and generally not flat in the other.
By assuming that we can treat data as realizations of parameters, we are incorporating the principle of Bayesian updating into our description of probability densities that can be applied in the same way to the prior, likelihood and posterior distributions. 

Global flatness in the Gaussian abstract space is also helpful in practice, as it allows us to perform operations in flat space and map the results back to the original parameter space of interest. This is a consequence of the fact that  
the reparameterization to abstract parameter space using normalizing flows is an isometry.
\begin{figure}
\includegraphics[width=\columnwidth]{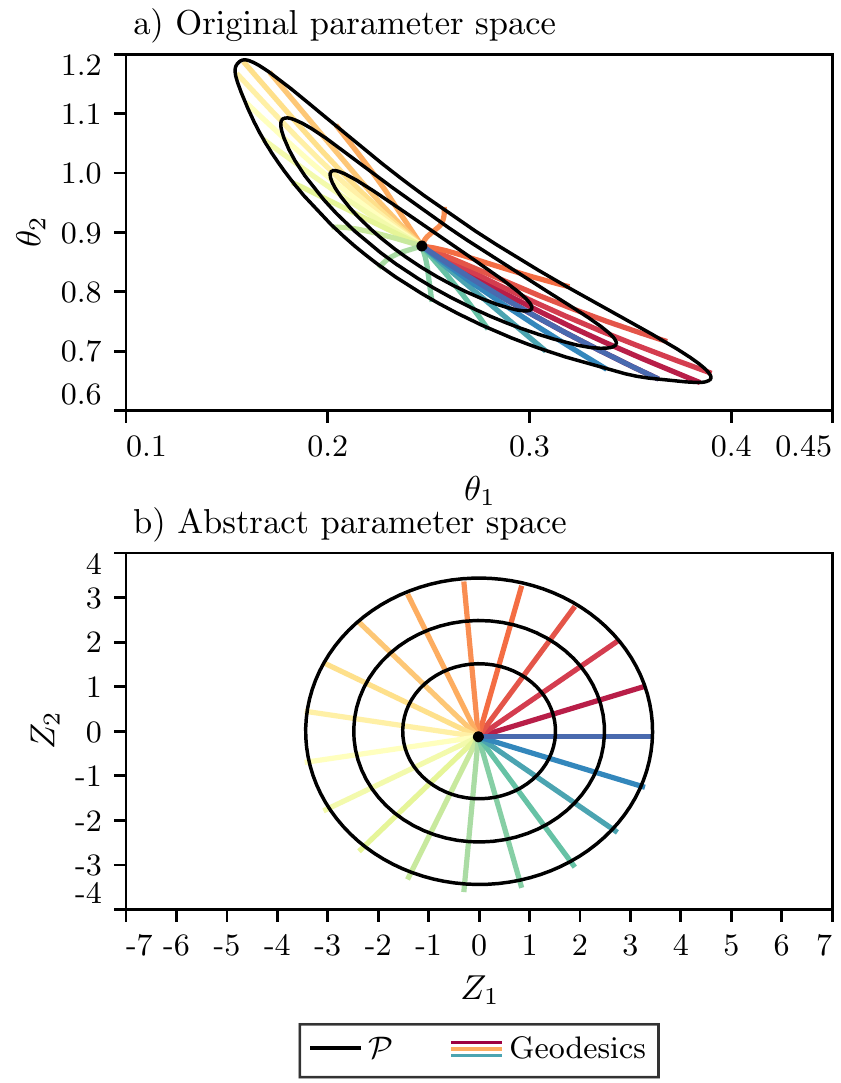}
\caption{ \label{fig:geodesics}
Geodesics of the metric in parameter space shown in the original and abstract parameter space, in the upper and lower panels  respectively. 
Straight line geodesics computed at evenly spaced angles around the maximum a posteriori probability (MAP) estimate in abstract space are transformed to original space.
In both panels black lines indicate the 68\%, 95\% and 99.7\% C.L. regions of the example probability density. 
}
\end{figure}
In \cref{fig:geodesics}, we illustrate an example of this property with the relationship between geodesics of the metric in abstract and original parameter spaces. 
In the Gaussian abstract space, geodesics are straight lines, and thus trivially solved for.
Once we have the geodesics of interest, we can map them to the original space, where the problem would be significantly more complicated to solve, due to the fact that the connection does not vanish.

Geodesic length is defined as the distance between any two points along the geodesic that connects them. Note here that geodesic length is invariant under reparameterizations, such as $\phi$. The length of the red line in panel b) is the same as the length of the red line in panel a) defined with respect to the metric $\mathcal{F}$. This is because reparameterizations are isometries that preserve the infinitessimal line element and angles between two corresponding sets of vectors.

\subsection{Local Principal Component Analysis} \label{Sec:LPCA}
Once we have a metric at every point in parameter space, we can promote the global version of the PCA problem, that we discussed in previous sections, to a local version:
\begin{align} \label{eq:LPCA}
\mathcal{F}_{\mu\nu}(\theta) u^\nu = \alpha \delta_{\mu\nu} u^\nu  \,,
\end{align}
subject to the constraint:
\begin{align}
u^\mu \delta_{\mu\nu} u^\nu = 1 \,.
\end{align}
Differently from the previous sections, here the PCA problem is assumed to be dependent on the position in parameter space, since $\mathcal{F}_{\mu\nu}$ depends on it. We therefore call it Local PCA hereafter.
This form of the eigenvalue problem makes it immediately clear that this is not a covariant problem since $\delta_{\mu\nu}$ is not a tensor and the norm of the eigenvector is computed with the Euclidean metric and would change when we change coordinates.

The Local Principal Components (PCs) can be understood in much the same way as the usual PCA. 
At any given point they are the orthonormal directions in which the local variance is maximized. 
If the original parameter space already has a Gaussian posterior distribution, $\mathcal{F}_{\mu\nu}$ would be constant and the results of the Local PCA would be uniform everywhere and agree with global PCA.
However, this is not in general the case with non-Gaussian distributions. 

\begin{figure}% [!ht]
\centering
\includegraphics[width=\columnwidth]{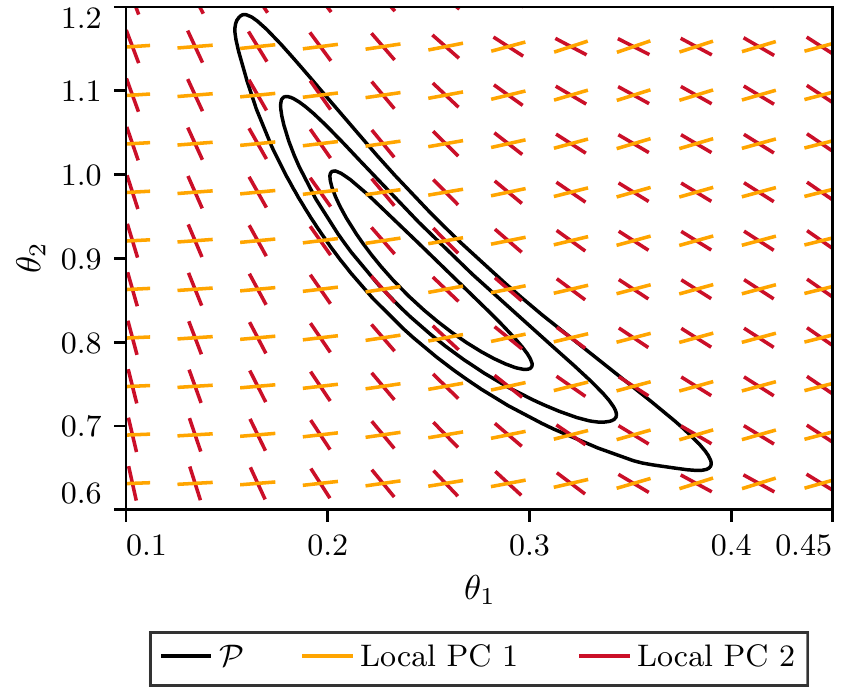}
\caption{ \label{fig:example_2_PCA_coords}
Local Principal Components computed on a grid in parameter space for a test example.
Different colors show different modes, the black lines indicate the 68\%, 95\% and 99.7\% C.L. regions of the example probability density.
}
\end{figure}

We see in \cref{fig:example_2_PCA_coords} an example of local PCs computed on a grid in parameter space for the example case we consider throughout this section.
As we can see the non-Gaussianity of the posterior translates in local rotations of the PC, especially when we compare the lower right corner to the upper left one.

\subsection{Principal Component Curves} \label{sec:PCC}

It is natural to trace the Local PCs to obtain PCA curves, that we denote with $\gamma(\lambda)$, and that we refer to as Principal Component Curves (PCC).

Along these, we would continuously be diagonalizing the local metric.
Since the Local PCs are orthonormal (for the metric $\mathcal{F}$), PCCs crossing at any given point would be either parallel or perpendicular to each other at this point. Correspondingly, the parameters along PCCs would be locally statistically independent.

The natural geometric way of finding such curves is to define the vectors $u^\mu$ that parallel transport the local PCA constraint equation:
\begin{equation} \label{Eq:LPCAtransport}
u^\alpha \nabla_\alpha f_{\mu}
= u^\alpha \partial_\alpha f_{\mu}
- u^\alpha \Gamma^{\lambda}_{\alpha\mu}f_{\lambda} = 0,
\end{equation}
where ${f_{\mu}\equiv\mathcal{F}_{\mu\nu} u^\nu - \alpha \delta_{\mu\nu} u^\nu}$ and $\Gamma^{\lambda}_{\alpha \mu}$ is the Christoffel symbol describing the metric connection associated with $\mathcal{F}_{\mu\nu}$. Then, we define the curves $\gamma^\mu$ based on the vectors $u^\mu$ defined above as:
\begin{align} \label{Eq:LPCAposition}
\frac{d \gamma^\mu}{d\lambda} = u^\mu \,.
\end{align}
Since the normalizing flow transformation is invertible, the metric tensor field is continuous and is non-singluar and this equation has unique solutions.
We note that since the initial velocity vector has to satisfy the Local PCA constraint, only a discrete set of initial velocities is admissible at any given point.
We chose to parametrize the curve with geodesic length:
\begin{align}
\dd{s} = \left( \mathcal{F}_{\mu\nu} u^\mu u^\nu \right)^{1/2} \dd{\lambda} \,,
\end{align}
so that the total length of a given curve is parameter invariant.

\begin{figure}% [!ht]
\centering
\includegraphics[width=\columnwidth]{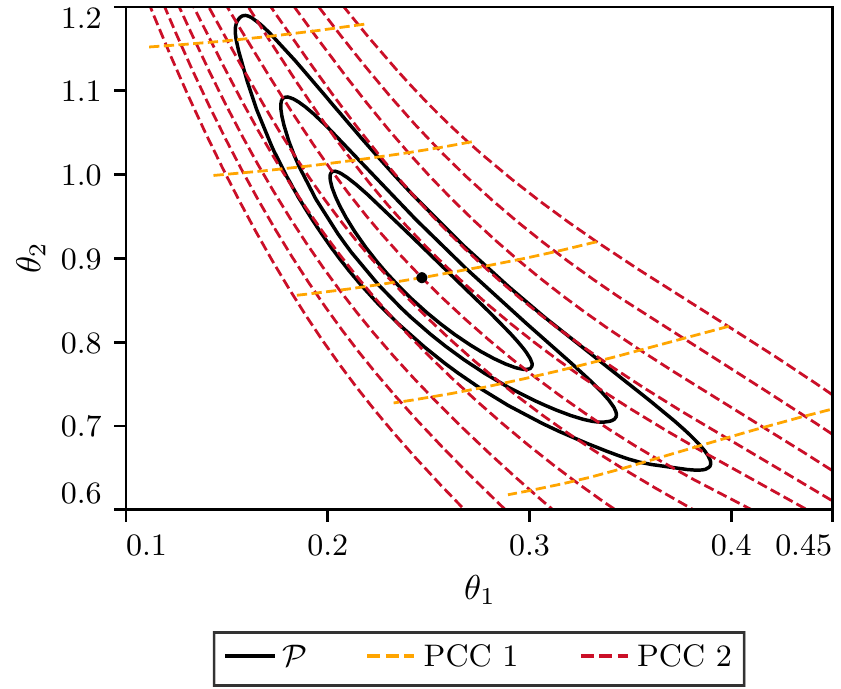}
\caption{ \label{fig:example_2_PCA}
Principal Component Curves (PCCs) computed for the test example.
Different colors show different modes, the black lines indicate the 68\%, 95\% and 99.7\% C.L. regions of the example probability density.
}
\end{figure}
We see in \cref{fig:example_2_PCA} the curves of the two modes for the example model we consider.
As we can see these are not straight lines, corresponding to the changes in directions of the local PCs shown in \cref{fig:example_2_PCA_coords}.

Note that we cannot simply integrate \cref{Eq:LPCAposition}, for a given mode, because, locally, eigenvalues might become degenerate along the flow, i.e. at a given point two eigenvalues could be the same and we would loose uniqueness, and therefore the direction to take. The key is to find the gradient of the eigenvector and solve the second order differential equation.
Solving the transport equation for the Local PCA problem ensures that we can follow continuously a given trajectory and integrate through points of local eigenvalue degeneracy.
On the other hand, local eigenvectors can never be degenerate since the Jacobian of the normalizing flow is invertible and the metric has always full rank.

\Cref{Eq:LPCAtransport} is impractical to solve since it contains connection terms which are expensive to compute numerically.  
For this reason, it is much easier to solve the Local PCA transport equation in abstract coordinates, where the connection vanishes since the space is flat. Note that while parameters transform as $\theta \mapsto \phi(\theta)$, vector fields transform as $ u(\theta) \mapsto  J_\phi u(\theta)$

For our PCCs, we define $u^\mu$ as the vector field that diagonalizes $\mathcal{F}=J_{\phi}^TJ_{\phi}$, so that \cref{eq:LPCA} becomes:
\begin{align}
J_{\phi}^TJ_{\phi} u &= \alpha u,
\end{align}  
which, in abstract space, reads:
\begin{align} \label{eq:diagineuc}
     J_\phi J_{\phi}^T w &= \alpha w
\end{align}   
where $w$ is now a vector field in abstract space corresponding to $u$, that diagonalizes $J_{\phi}J_\phi^T$.
Therefore, instead of solving for $\gamma$ such that $\dot{\gamma} = u$ and $\mathcal{F}u = \alpha u$ in the original parameter space, we can solve for $\gamma^\prime$ such that $\dot{\gamma}^\prime = J_\phi\ \dot{\gamma} = J_\phi \ u \equiv w, \ J_\phi J_\phi^T w = \alpha w$, and map the curve back to the original parameter space. The desired normalization for $w$ can be derived from that for $u$:  \begin{align}
   1= u^\mu u^\nu \delta_{\mu\nu}&= w^T J_\phi^{-T} J_\phi^{-1}w = w^T \alpha^{-1}w, \\
   w^Tw &= \alpha 
\end{align}

Due to the possibility of degenerate eigenvalues, one cannot simply follow a given eigenvalue. This may result in non-smooth and ill-behaved curves. To solve this issue, we can enforce a smooth evolution of $\alpha(t)$ and $w(t)$ with a differential equation to ensure that we are tracing the same degree of freedom at each point in parameter space, even if the ordering of the eigenvalues changes. Starting from the time derivative of $w^TJJ^Tw = \alpha^2$ and ${w^T\dot{w}=\dot{\alpha}/2}$ (and using $J$ as shorthand for $J_{\phi}$), we have:
\begin{align}
    2\dot{\alpha}\alpha &= w^T(\dot{J}J^T + J\dot{J}^T)w + 2 w^T JJ^T \dot{w} \nonumber\\
    &= 2w^TJ\dot{J}^Tw + \alpha  \dot{\alpha}, \\
    \dot{\alpha} &= 2 w^TJ (\partial_w J)^Tw/\alpha \label{eq:alphadot}
\end{align}
where $\partial_w$ is the directional derivative in the $w=\dot{\gamma}^\prime$ direction. 
Given that diagonalization is a costly operation to perform at every point, we may wish to solve for $w$ from a differential equation. Starting from the time derivative of $JJ^Tw = \alpha w$, we further find:
\begin{align}
    (\dot{J}J^T+J\dot{J}^T)w + JJ^T\dot{w} = & \dot{\alpha}w+\alpha \dot{w}, \\
    \begin{split}
        [JJ^T- (w^TJJ^Tw)\mathds{1}]\dot{w} = & \left[2 (w^TJ (\partial_w J^T)w)/\alpha\mathds{1} \right. \\
        &  \left. -(\partial_w J)J^T - J(\partial_w J)^T \right] w.
    \end{split}
      \label{eq:wdot}
\end{align}
The above is a second order differential equation entirely in terms of $\gamma^\prime(t)=w$. It assumes \cref{eq:alphadot}, which means that following $w$ this way solves the eigenvalue degeneracy problem as well as removing the need to diagonalize the metric at every point. Note that $JJ^T-\alpha \mathds{1}$ is non invertible as $\ker (JJ^T-\alpha\mathds{1}) = {\rm span}(w)$. However, with the constraint that $\dot{w}^Tw=\dot{\alpha}/2$, we can solve the above equation.
Though equation \cref{eq:wdot} does not contain connection terms, it is still difficult to compute in practice since it contains metric derivative terms. If there is reason to believe that the metric is never degenerate in the region of interest then a simpler problem can be solved by just diagonalizing $JJ^T$ at each point and following the PCC in abstract space.

\begin{figure}
\centering
\includegraphics[width=\columnwidth]{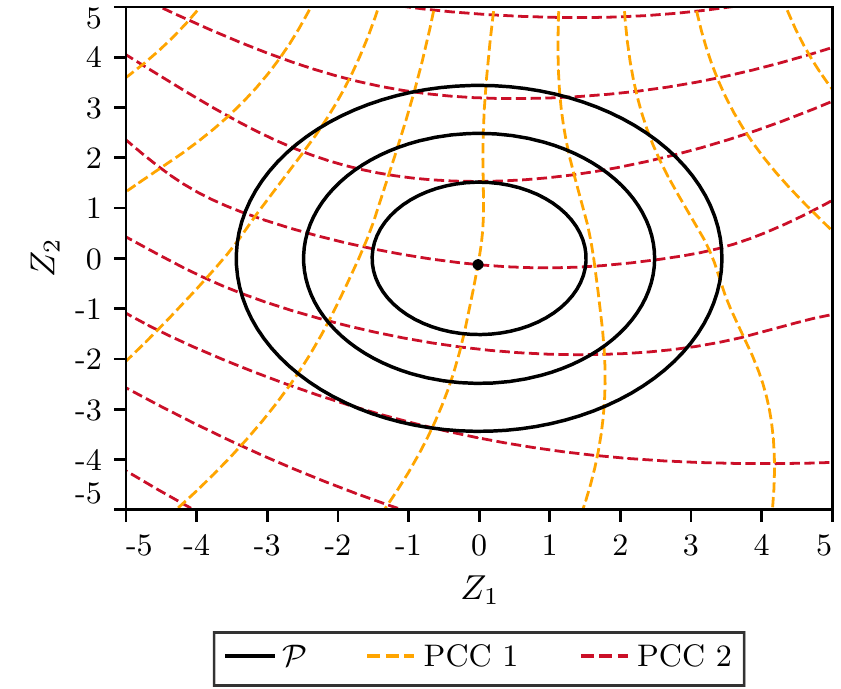}
\caption{ \label{fig:example_2_PCA_abs}
Principal Component Curves (PCCs) as in \cref{fig:example_2_PCA}, computed for the test example in abstract space.
Different colors show different modes, the black lines indicate the 68\%, 95\% and 99.7\% C.L. regions of the example probability density.
}
\end{figure}

\Cref{fig:example_2_PCA_abs} shows the PCCs produced in gaussianized coordinate space. These PCCs correspond to those shown in \cref{fig:example_2_PCA}, where the relationship between the PCCs and the posterior were evident. That is not the case here, as the vector fields generating these PCCs relate to the metric in  \cref{fig:example_2_PCA}, rather than the Euclidean metric. 

\begin{figure}
\centering
\includegraphics[width=\columnwidth]{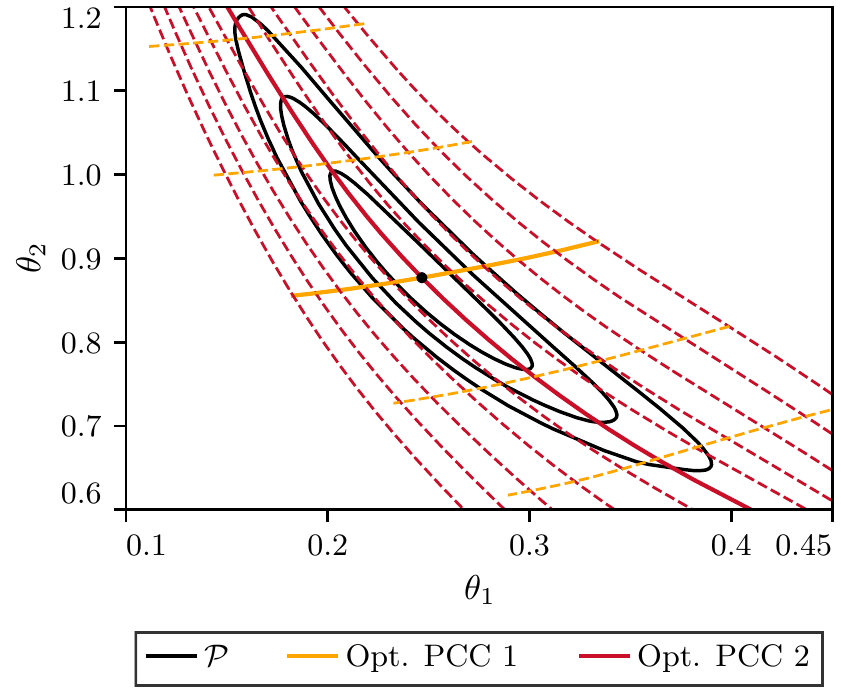}
\caption{ \label{fig:example_2_PCA_lines}
Principal Component Curves (PCCs) computed for the test example along with the Optimal PCCs obtained by minimizing the expectation of perpendicular distance.
Different colors show different modes, the black lines indicate the 68\%, 95\% and 99.7\% C.L. regions of the example probability density.
}
\end{figure}
\subsection{Perpendicular Distance} \label{sec:PerpDist}
In \cref{Eq:PlaneKL}, we noted that for every CPCA there exists a hyper-plane that runs perpendicular to it. This was useful in computing the contribution of a parameter to the variance of a given CPC mode. In a similar sense, given a family of PCCs there exists a hyper-surface that at every point intersects with one of the PCCs orthogonally. 

All curves on the hyper-surface will be orthogonal to the given family of PCCs. Therefore, minimizing distance on the hyper-surface is equivalent to the problem of finding a path of minimal distance while travelling strictly perpendicular to a given unit vector field, specifically the vector field defining the family of PCCs. Suppose $(M_1,\mathcal{F}_1)$ is a manifold and $u$ some vector field. We wish to find a curve $g(t)$ on $M_1$ that minimizes distance while remaining perpendicular to some vector field with respect to the $\mathcal{F}_1$ metric: 
\begin{align}
    &\mbox{maximize} \enspace  \int_I \sqrt{\dot{g}(t)^T \mathcal{F}_1 \dot{g}(t)} \dd{t} \nonumber\\
    &\mbox{subject to}  \enspace   \dot{g}(t)^T \mathcal{F}_1 u(g(t)) = 0 \label{eq:orthomin1}
\end{align}
Suppose now that we have some other manifold $(M_2,\mathcal{F}_2)$ and that we can map between the two manifolds with an isometry $\phi$. Let $g^\prime(t) = \phi(g(t))$ define the corresponding curve on $M_2$, and let $w=J_\phi\ u$ define the corresponding vector field on $M_2$. Now, note that since $\phi$ is an isometry:
\begin{align}
    \dot{g}(t)^T \mathcal{F}_1 \dot{g}(t) &= (J_\phi \ \dot{g}(t))^T \mathcal{F}_2 (J_\phi \ \dot{g}(t)) \nonumber \\
    &= \dot{g}^\prime(t)^{ T}\mathcal{F}_2 \dot{g}^\prime(t)  \\
    \dot{g}(t)^T \mathcal{F}_1 u(g(t)) &= (J_\phi \ \dot{g}(t))^T \mathcal{F}_2 (J_\phi \ u(g(t))) \nonumber \\
    &= \dot{g}^\prime(t)^{ T}\mathcal{F}_2 w(g^\prime(t))  
\end{align}
Additionally, given that $\phi$ is a bijection, picking a $g(t)$ is equivalent to just picking a $g^\prime(t)$. With all that, we see that the above optimization problem is equivalent to:
\begin{align}
    &\mbox{maximize} \enspace  \int_I \sqrt{\dot{g}^\prime(t)^{ T}\mathcal{F}_2 \dot{g}^\prime(t)} \dd{t} \nonumber\\
    &\mbox{subject to}  \enspace   \dot{g}^\prime(t)^{ T} \mathcal{F}_2 w(g^\prime(t)) = 0 \label{eq:orthomin2}
\end{align}
That is, solving the first optimization problem is equivalent to solving the second and mapping back with $\phi^{-1}$.

We therefore opt to solve this problem in the Gaussian abstract space, where $\phi$ is the normalizing flow mapping, such that we can solve the problem with a trivial metric and map it back to the original case of interest. Let $u$ be some vector field on the abstract space and $g(t)$ a curve on the same space that minimizes distance while remaining perpendicular to $u$. This is a constrained optimization problem, whose constraint one may naively think to enforce with a Lagrangian multiplier. However, Lagrangian multipliers are best applied to holonomic constraints, or constraints expressible with positions ($g(t)$) and not velocities ($\dot{g}(t)$). The constraint here is $\dot{g}(t) \cdot u(g(t))=0$, which is not holonomic. Fortunately, non-holonomic constraints linear in components of the velocity can be applied with Lagrange multipliers following the prescription in~\cite{holonomy}. The Lagrangian is:
\begin{align}
    L = \frac12 \dot{g} \cdot \dot{g} + \lambda \dot{g} \cdot u(g(t))
\end{align}
where $\lambda$ is the Lagrangian multiplier and the resulting equations are:
\begin{align}
    \ddot{g} + \lambda u &= 0, \quad\quad
    \dot{g} \cdot u = 0 .
\end{align}
Without solving for $\lambda$, the first equation makes sense heuristically: the velocity vector only needs to change in response to changes in the perpendicular vector field. 
Taking the time derivative of the second equation and taking the inner product of the first equation with $u$ we get:
\begin{align}
    \ddot{g} \cdot u + \dot{g} \cdot (\partial_{\dot{g}})u &= 0, \quad    u \cdot( \ddot{g} + \lambda u )= 0\\
    &\Rightarrow \dot{g} \cdot \partial_{\dot{g}} u = \lambda,
\end{align}
where $\partial_v$ is the directional derivative in the $v$ direction. The equation of motion to follow is given by:
\begin{align}
    \ddot{g} = -( \dot{g} \cdot \partial_{\dot{g}} u ) u. \label{eq:orthopath}
\end{align}
Given any initial position and velocity, solving this equation produces a curve that minimizes distance while remaining perpendicular to $u$. In the case where $u$ is a constant vector field, $\ddot{g}=0$ as expected. Along the solution, $\dot{g} \cdot u$ remains constant:
\begin{align}
    d_t( \dot{g} \cdot u)&= \ddot{g} \cdot u + \dot{g} \cdot (\dot{g}\cdot\partial) u \nonumber \\
    &=  -\dot{g} \cdot \partial_{\dot{g}} u  + \dot{g} \cdot \partial_{\dot{g}}u =0
\end{align}
So if we start with $\dot{g}^\prime \perp w$ it remains the case all the way through. In this way we can compute the ``orthogonal distance'' between two points with respect to some vector field. From this we can also define the orthogonal distance between a point and a curve as the shortest orthogonal distance between the point and any point on the curve. 
\subsection{Optimal Principal Component Curves} \label{sec:OptimalPCC}
Given the family of PCCs we can define the curve that best represents the constrained direction in parameter space by minimizing the expectation value between points in the parameter space and the curve, along the other PCC directions. 
We call the resulting curves the ``Optimal PCC''. 

However, the ``distance'' considered in the expectation value is not simply  the minimal distance between the curve and the point. It should be the minimal distance between the point and the curve while travelling strictly perpendicular to the family of PCCs. This distance would quantify the distance between point and curve strictly independent to the parameter defining the family of PCCs. The method of computing this notion of perpendicular direction was discussed in the previous section.

Suppose again we have a statistical manifold $(M_1, \mathcal{F}_1)$ and a probability distribution $P_1$ on $M_1$. Let $G = \{ \gamma_i \}_i$ be a family of parallel PCCs on $M_1$ deriving from varying the same parameter. Let $u$ be the vector field such that at $p\in \gamma_k$, $u=\dot{\gamma}_k$. Let $d(x, y,u)$ be the distance between $x,y\in M_1$ on the shortest path that remains orthogonal to vector field $u$. Define $d(x,\gamma(t),u) \equiv \min_{y\in \gamma} d(x,y,u)$ where $\gamma$ is some curve. We can find the Optimal PCC by solving this optimization problem:
\begin{align}
        \min_{\gamma\in G} \mathbb{E}[d(x,\gamma,u) ]  &= \min_{\gamma\in G} \int_{M_1} d(x,\gamma,u)  P_1(x) \dd{x} 
\end{align}

We can introduce an arbitrary reparameterization $\phi$ and derive a new statistical manifold  $(M_2, \mathcal{F}_2)$ with a probability distribution $P_2$ on $M_2$. The probability distribution transforms as $P_2 = P_1/\abs{J_\phi}$. As seen previously, reparameterization induces an isometry between the two statistical manifolds. Following the argument above that \cref{eq:orthomin1} and \cref{eq:orthomin2} are equivalent, $d(x,y,u) = d(\phi(x), \phi(y), J_\phi \ u)$. Since $\phi$ preserves values of $d(x,y,u)$, its minimum among $y\in \gamma$ must also be the same across $\phi$. That is, $d(x,\gamma,u)= d(\phi(x), \phi(\gamma),J_\phi \ u)$.

Now, note that:
\begin{align}
    \mathbb{E}&[d(x,\gamma,u) ] \nonumber = \int_{M_1} d(x,\gamma,u)  P_1(x) \dd{x} \nonumber\\
    &= \int_{M_1} d(\phi(x),\phi(\gamma),J_\phi \ u ) P_2(\phi(x)) \abs{J_\phi} \dd{x} \nonumber\\
    &= \int_{M_2} d(\phi(x),\phi(\gamma),J_\phi \ u ) P_2(\phi(x)) \dd(\phi(x)) \nonumber\\
    &= \mathbb{E}[d(\phi(x),\phi(\gamma),J_\phi \ u )]
\end{align}

Now, suppose that $\phi(G)=\{\phi(\gamma_i)\}_i$ the family of PCCs corresponding to $G$ on $M_2$. If the solution to the optimization problem
\begin{align}
    \min_{\gamma\in G} \mathbb{E}[d(x,\gamma,u) ]
\end{align}
is $\gamma_i$ then the solution to the optimization problem 
\begin{align}
    \min_{ \xi \in \phi( G)} \mathbb{E}[d(\phi(x), \xi, J_\phi \ u)] \label{eq:optimizationforprimaryflows}
\end{align}
is $\phi(\gamma_i)$. That is, the Optimal PCCs are generally covariant. By this we mean that every step of the procedure to find the PCC from the local PC vector field is generally covariant. 
Finding a PCC in one parameterization gives the same result as finding a PCC in a different parameterization and mapping it back. The minimal distance between two points along a path orthogonal to a vector field is reparameterization invariant. The optimization problem to find the Optimal PCCs is also reparameterization invariant. So, once again, we can opt to solve this optimization problem in the abstract space, letting $\phi$ be the normalizing flow transformation that Gaussianizes the posterior. 

We show the Optimal PCC for our test example in \cref{fig:example_2_PCA_lines}.
The first Optimal PCC is the curve, in the first family of PCCs, with the smallest expectation value of distance between that curve and points in the parameter space, along the direction of the second family of PCCs.
The same applies to the second Optimal PCC. 
In the example case in \cref{fig:example_2_PCA_lines} Optimal PCCs appear to cross the MAP estimate, but this is not necessarily always the case.

\begin{figure}
\centering
\includegraphics[width=\columnwidth]{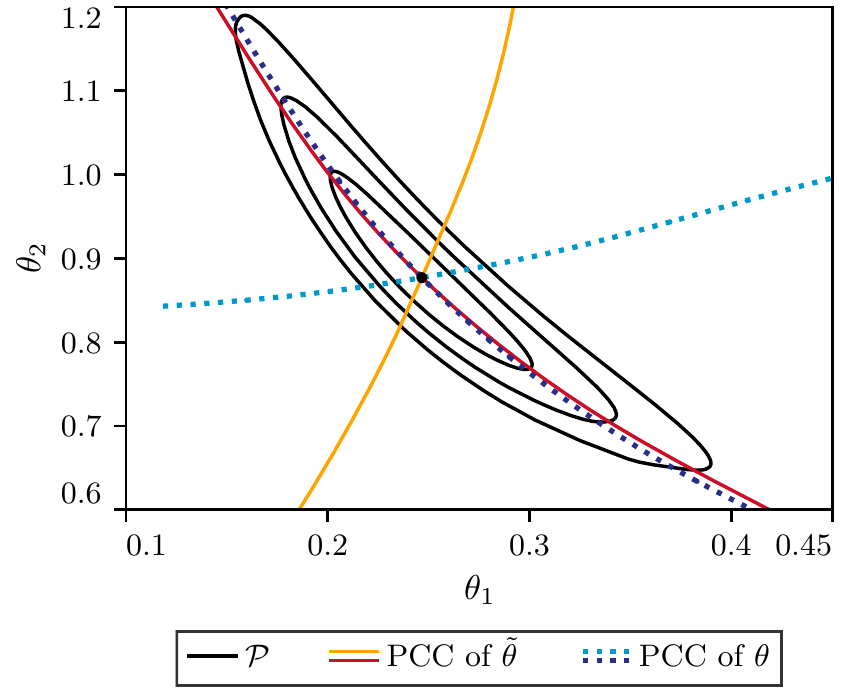}
\caption{ \label{fig:example_2_LPCA_non_covariance}
An example of how Principal Component Curves (PCCs) differ depending on the parameter basis that is chosen. The PCCs are calculated in the original parameter basis $\theta$ and in a non-linearly transformed basis 
$\tilde{\theta}$ and transformed back, resulting in distinct curves.
Black lines indicate the 68\%, 95\% and 99.7\% C.L. regions of the example probability density.
}
\end{figure}

The most fundamental step in obtaining Optimal PCCs, however, in \cref{eq:LPCA}, is not reparametrization invariant, making the PCCs not generally covariant. 
In \cref{fig:example_2_LPCA_non_covariance}, we show the Optimal Principal Component Curves calculated in the original parameter basis and also in a reparametrized basis. We use as a transformation the logarithm of both parameters, that for our test example makes the distribution globally Gaussian.
Calculating the optimal curves in the two bases and mapping the results gives visibly distinct curves, confirming the lack of covariance of Principal Component Curves. 

\subsection{Local Covariant Principal Component Analysis} \label{Sec:LCPCA}
We can solve this problem of lack of covariance analogously to what we have shown in the linear model in \cref{Sec:KL}, by considering the local Covariant PCA (CPCA) of the Fisher matrix with respect to some reference tensor:
\begin{align}
\mathcal{F}_{\mu\nu}^{p} u^\nu = \alpha \mathcal{F}_{\mu\nu}^{\Pi} u^\nu \label{eq:LKL}
\end{align}
where $\mathcal{F}_{\mu\nu}^{p}$ is the Fisher matrix discussed in the previous section and $\mathcal{F}_{\mu\nu}^{\Pi}$ is the reference tensor we derive from the prior distribution. We use normalizing flows to gaussianize the \textit{prior} distribution, then assuming that the tensor $\mathcal{F}_{\mu\nu}^{\Pi} = \delta_{\mu\nu}$ when the distribution is a unit gaussian, we derive $\mathcal{F}_{\mu\nu}^{\Pi}$ using the normalizing flows. This is the same procedure with which we derive $\mathcal{F}_{\mu\nu}^{p}$ so $\mathcal{F}_{\mu\nu}^{\Pi}$ may be interpreted as the local Fisher matrix for the prior.

Just as linear CPCA was affine invariant, local CPCA is generally covariant.
This is not entirely surprising as reparameterizations locally manifest as affine transformations. 
Therefore, it is intuitive that reparameterization preserves local CPCA but not local PCA, just as affine transformations had preserved CPCs but not PCs. 

One can generate vector fields of local CPCA components and follow the vector fields to get Covariant Principal Component Curves (CPCCs) that locally solve the CPCA problem. One can also define the ``Optimal CPCC'' as the CPCA equivalent to the Optimal PCC. 
The end result would now be fully generally covariant. 

The only difference in generating the right vector field in abstract space where we prefer to work as the metric is trivial. 
Let $\phi$ be the normalizing flow transformation that gaussianizes the posterior. Then, in the abstract space we need to solve for $w = J_{\phi} u$. Starting from \cref{eq:LKL} and recalling that $\mathcal{F}^{p} =J_{\phi}^TJ_{\phi}$:
\begin{align}
    \mathcal{F}^{p} J_{\phi}^{-1} w   &= \alpha \mathcal{F}^{\Pi} J_{\phi}^{-1} w  \\
     w   &= \alpha J_{\phi}^{-T} \mathcal{F}^{\Pi} J_{\phi}^{-1} w 
\end{align}
instead of \cref{eq:diagineuc}. Therefore, the correct vector field in the abstract space is one that diagonalizes $J_\phi^{-T}\mathcal{F}^{\Pi}  J_\phi^{-1}$. Note that this is consistent with the local PCA case when $\mathcal{F}^{\Pi} = I_n$.

\begin{figure}
\centering
\includegraphics[width=\columnwidth]{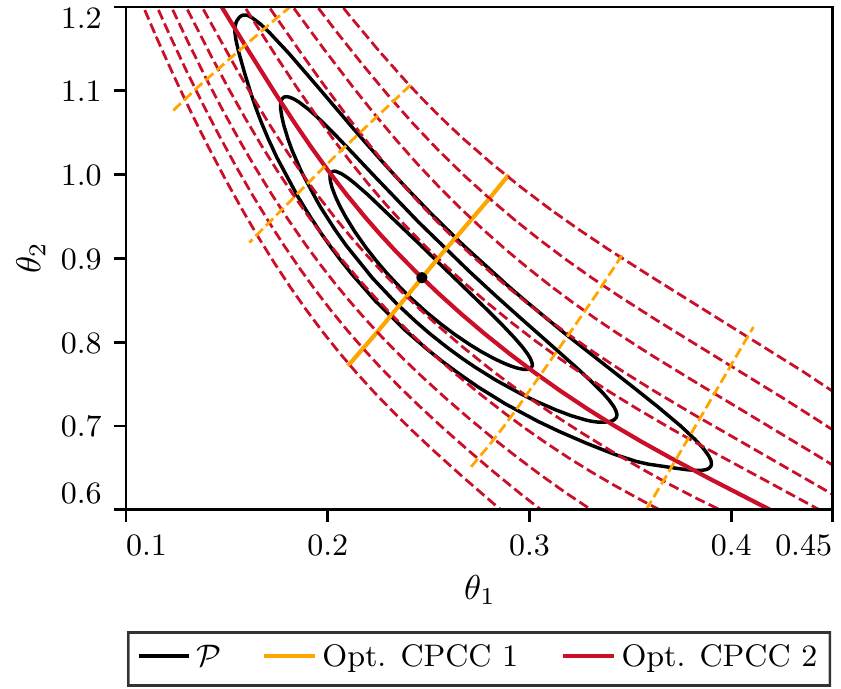}
\caption{ \label{fig:example_2_KL}
Covariant Principal Component Curves (CPCCs) computed for the test example along with the Optimal CPCCs.
Different colors show different modes, the black lines indicate the 68\%, 95\% and 99.7\% C.L. regions of the example probability density.
}
\end{figure}
We show in \cref{fig:example_2_KL} the CPCC analysis applied to the test example of this section.
As we can see the CPCCs are not Euclidean orthogonal but are orthogonal in the units of the posterior distribution, as a result of the CPCA constraint, \cref{Eq:DecorrelateKLFisher}, that still holds locally.
In this example the Optimal CPCCs are derived by minimizing the expectation of orthogonal distance.

\begin{figure}
\centering
\includegraphics[width=\columnwidth]{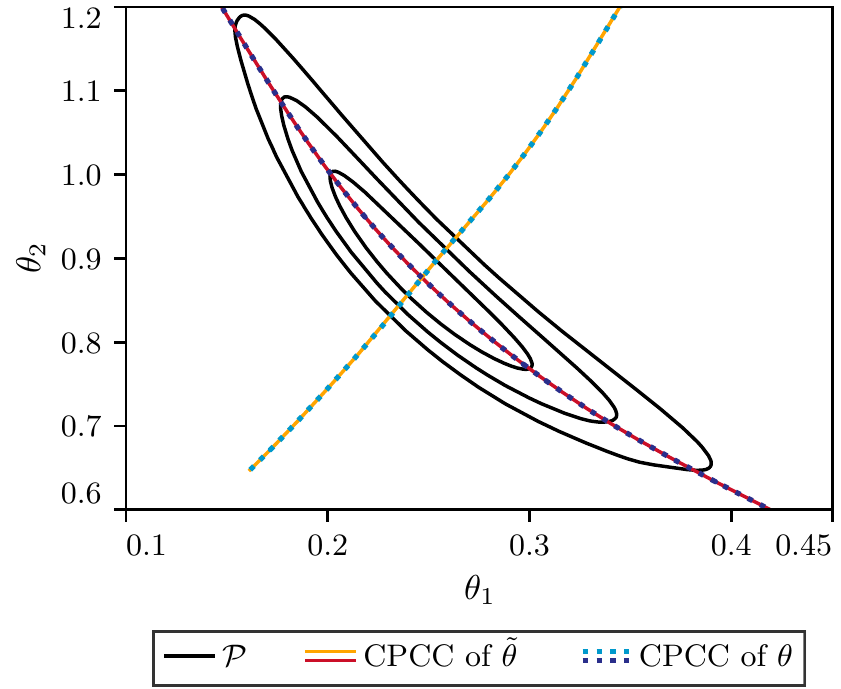}
\caption{ 
\label{fig:example_2_LCPC_covariance}
An example of how Covariant Principal Component Curves (CPCCs) are independent of the parameter basis that is chosen. CPCCs are calculated in the original parameter basis $\theta$ as well as in a non-linearly transformed basis 
$\tilde{\theta}$ and transformed back, resulting in identical curves.
Black lines indicate the 68\%, 95\% and 99.7\% C.L. regions of the example probability density.
}
\end{figure}

Analogously to \cref{fig:example_2_LPCA_non_covariance} for the PCC analysis, in \cref{fig:example_2_LCPC_covariance}, we show the Optimal CPCCs calculated in the original parameter basis and in a reparametrized basis taking the logarithm of parameters. Calculating the curves in the two bases results in identical curves, clearly showing the covariance of this method.

\section{Benchmark Examples} \label{Sec:ToyExamples}
In this section, we apply the results of the previous section to a set of benchmark examples.
We first consider, in \cref{Sec:ToyExample.StrongNonGaussian}, a strongly non-Gaussian example.
In \cref{Sec:ToyExample.PriorInformed}, we consider an example with an unconstrained direction that is prior informed.

\subsection{Strong non-Gaussianity} \label{Sec:ToyExample.StrongNonGaussian}
In the first example case we consider a posterior distribution that has a strong, banana-shaped  non-Gaussian degeneracy.
The density of this distribution, within the prior boundaries, is given by ${\mathcal{P}(\theta_1,\theta_2) \propto \mathcal{N}(\sqrt{\theta_1^2 + 20(2 \theta_1^2-\theta_2-1/2)^2}, 1/4)}$, up to an irrelevant normalization constant.
The prior is flat for both parameters in the interval $[-3, 3]$ and is chosen to be isotropic (the same in both parameters) and uninformative.
\begin{figure*}
\centering
\includegraphics[width=\textwidth]{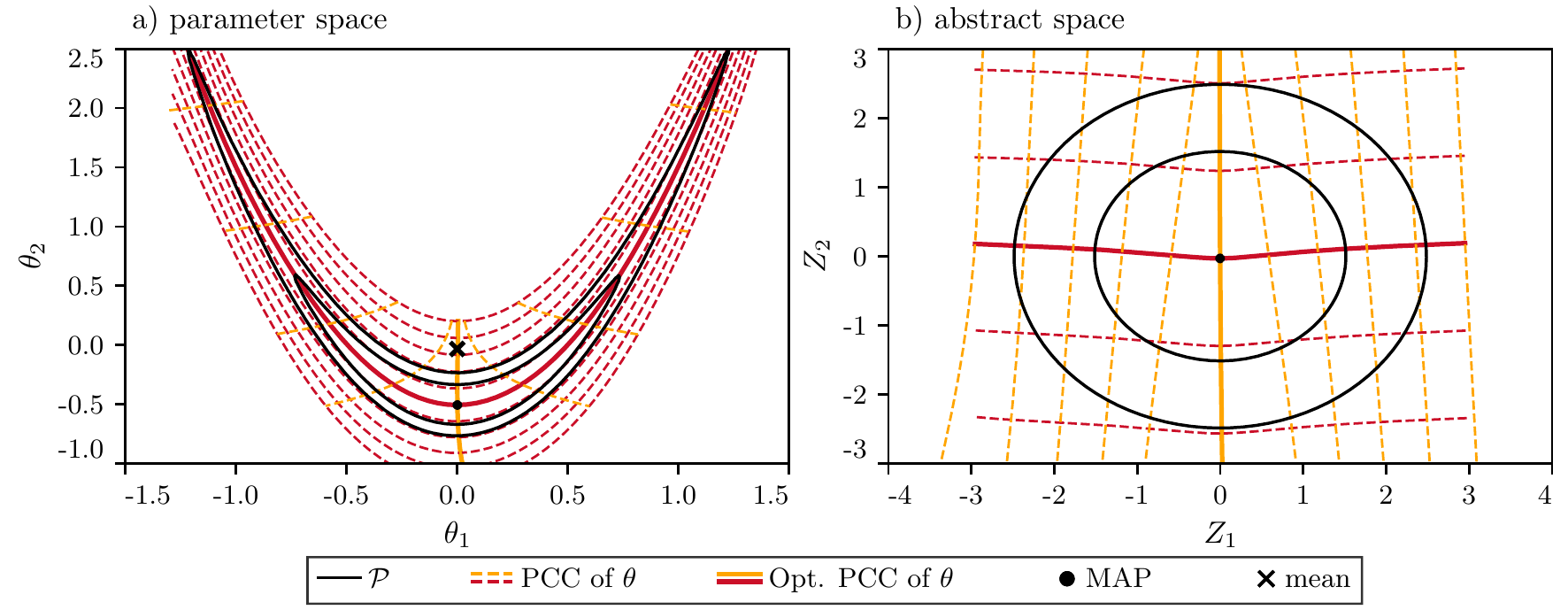}
\includegraphics[width=\textwidth]{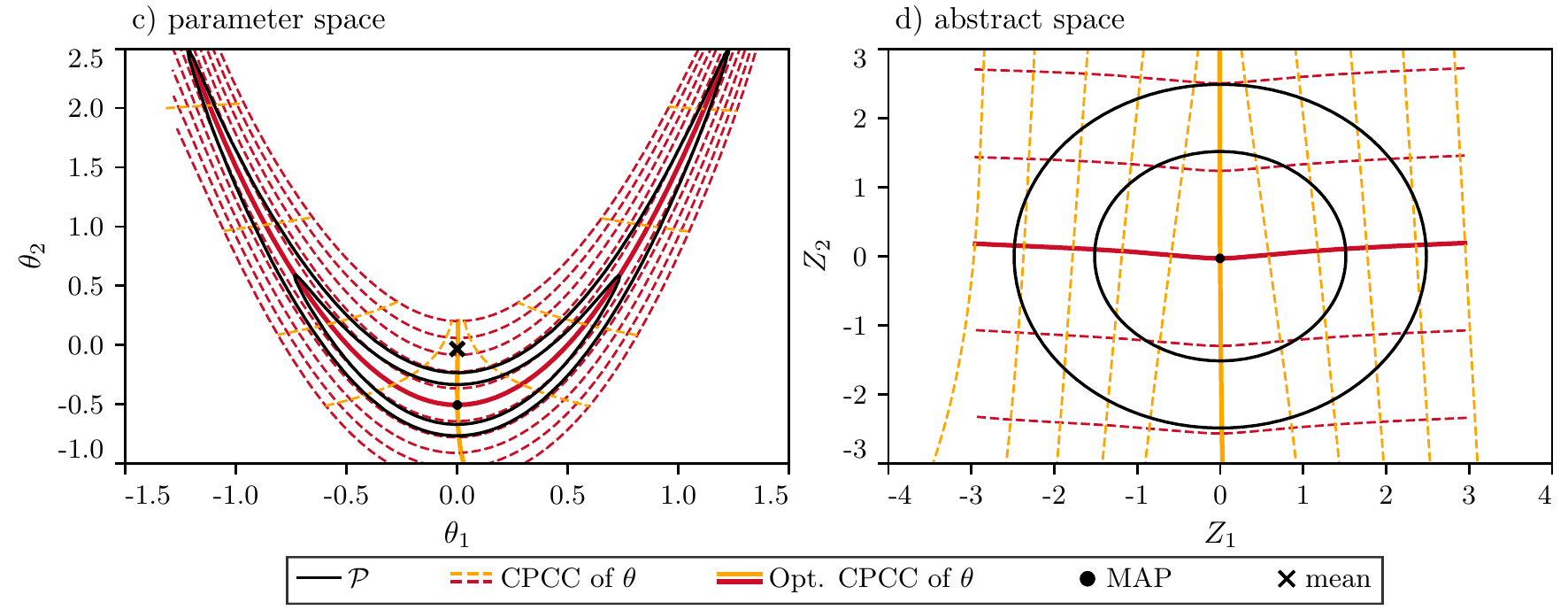}
\caption{ \label{fig:toy_example_3}
The families of Principal Component Curves (PCCs) and Covariant Principal Component Curves (CPCCs) for the test example discussed in \cref{Sec:ToyExample.StrongNonGaussian}.
Panel a) shows the two families of PCCs in parameter space while Panel b) shows them in abstract space where the posterior is Gaussian.
Panel c) shows the families of CPCCs while Panel d) shows them in abstract space.
Since the prior is isotropic, i.e. the same in both coordinates, PCCs and CPCCs coincide.
In all panels the black lines show the 68\% and 95\% C.L. region for the probability density we consider.
}
\end{figure*}
We can see in \cref{fig:toy_example_3}, Panels a) and b), the PCC analysis of this toy example.
As we can see both PCCs closely track the overall distribution and its full non-Gaussian structure of the posterior is resolved.
We can note some spurious curvature in the tails of the distribution that is due to the normalizing flow model lack of training samples in that region.
Optimal PCCs go through the MAP estimate and not the mean that, due to the strong non-Gaussianity, is located in a very low probability region.

In \cref{fig:toy_example_3}, Panels c) and d), we show the CPCC analysis.
The prior in this case is isotropic, so, up to a redefinition of the eigenvalues, the CPCCs should match the PCCs, as it clearly happens in this test case.
Note however that the two families of curves would not match if we were to change the parameter basis and work in a transformed basis.
In this case, the CPCCs would simply transform from the one shown in figure, while the PCCs could qualitatively change behavior.

\subsection{Informative prior} \label{Sec:ToyExample.PriorInformed}
\begin{figure*}
\centering
\includegraphics[width=\textwidth]{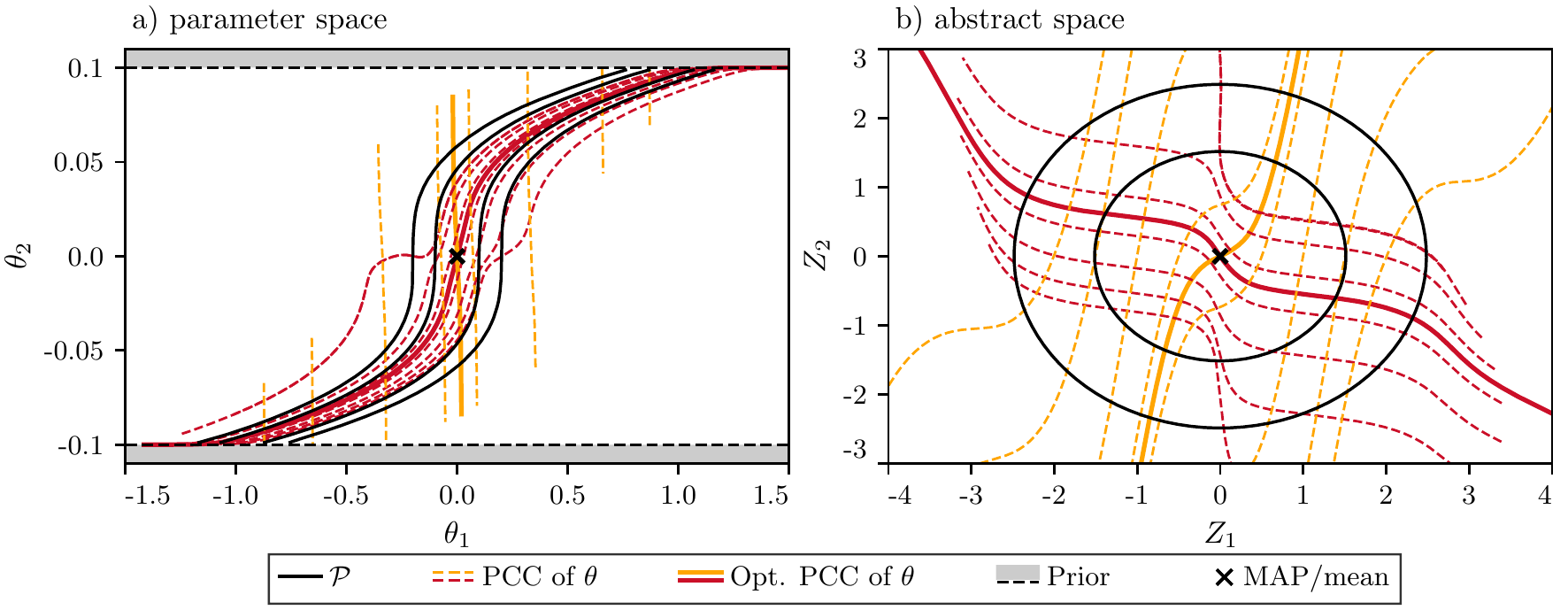}
\includegraphics[width=\textwidth]{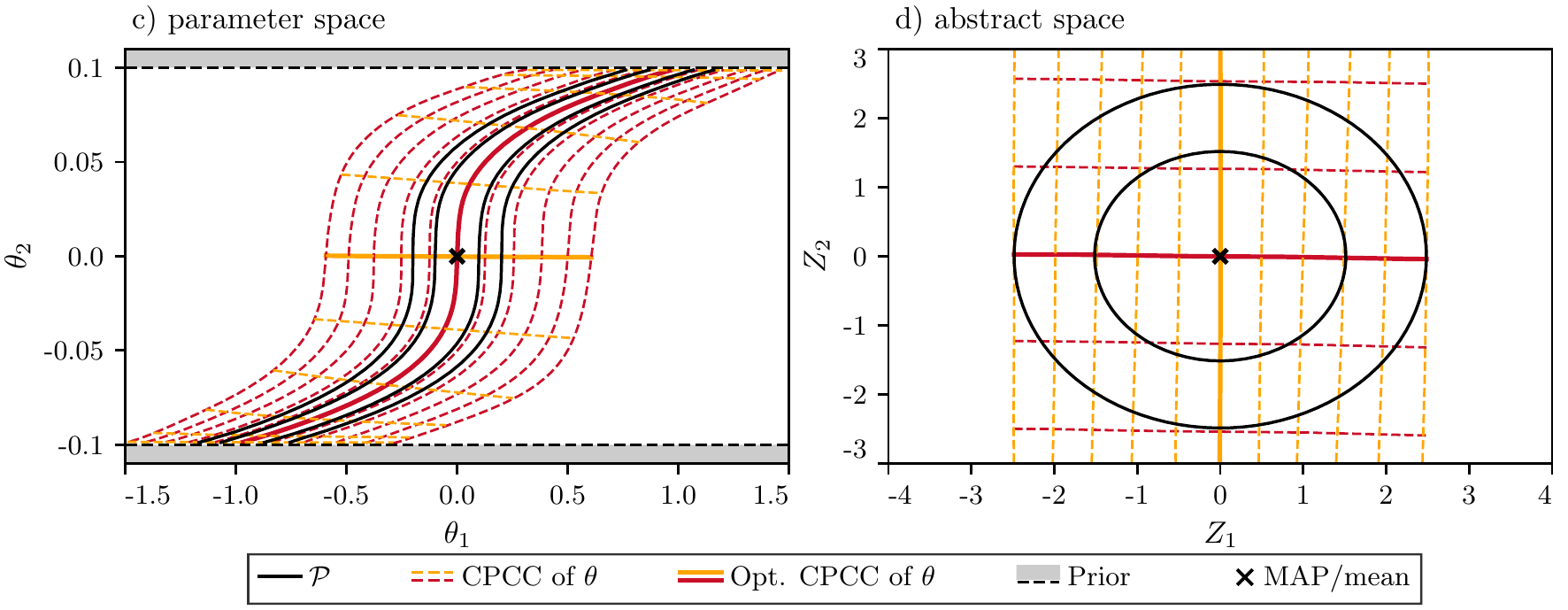}
\caption{ \label{fig:toy_example_4}
The families of Principal Component Curves (PCCs) and Covariant Principal Component Curves (CPCCs) for the test example discussed in \cref{Sec:ToyExample.PriorInformed}.
Panel a) shows the two families of PCCs in parameter space while Panel b) shows them in abstract space where the posterior is Gaussian.
Since the PCCs depends on the units of the parameters, that in this example are different by design, the two families of curves do not track the behavior of the posterior distribution well.
Panel c) shows the families of CPCCs while Panel d) shows them in abstract space.
The advantage of the CPCCs, with respect to the PCCs, is that they are fully covariant and, by taking into consideration the units of the problem, correctly track the probability density.
In all panels the black lines show the 68\% and 95\% C.L. region for the probability density we consider.
}
\end{figure*}

We now consider an example in which the prior is informative and the posterior distribution density is given by ${\mathcal{P}(\theta_1,\theta_2) \propto \mathcal{N}(\theta_1-(10 \theta_2)^3, 1/2)}$, up to an irrelevant normalization constant, within the prior boundaries.
The prior is chosen to be uniform and independent between the two parameters, in the intervals ${\theta_1 \in [-2, 2]}$ and ${\theta_2 \in [-0.1, 0.1]}$.
In this example, the prior is not isotropic, as it is usually the case when we deal with parameters that have different physical units. By construction, the probability distribution has a non-linear direction that is fully prior constrained.

We can see in \cref{fig:toy_example_4} the PCCs for this test example.
Differently from the previous example, the PCCs do not closely track the behavior of the probability density.
This happens because of its sensitivity to units and is reflected by the fact that the best constrained direction is aligned with the second parameter.
Within the PCC analysis, the second parameter appears much better constrained with respect to the first one because it varies on a much smaller scale, even though it is, by construction, almost totally unconstrained.
The second direction, shows, on the other hand, a strong non-linearity that is again an artifact of units.

We can compare the behavior of the CPCCs to the PCCs, shown in \cref{fig:toy_example_4}.
In this case, the CPCCs take into account the different units of the prior and follows correctly the behavior of the probability density.
In this case the best constrained direction tracks the decay of the probability density, while the second direction accurately tracks the non-linear degeneracy that is prior constrained.

\section{Application to cosmological data} \label{Sec:RealExamples}
In this section we discuss a series of examples from cosmology.

We use the measurements of galaxy clustering and lensing from the Dark Energy Survey (DES) and consider the publicly released measurements from the first year of data (Y1)~\cite{DES:2017myr}, in \cref{Sec:DES.shear} and \cref{Sec:DES.shear.to.3x2}. In \cref{Sec:DES.CMB.lensing}, we use CMB lensing reconstruction measurements from Planck~\cite{Planck:2018lbu} and compare them with DES measurements.

We consider the $\Lambda$CDM model with its five relevant cosmological parameters that are included in the analysis:
the total matter density parameter $\Omega_m$, including cold dark matter and baryons;
the baryon  density parameter $\Omega_b$; 
the value of the Hubble constant $H_0$ in km/s/Mpc; 
the rescaled amplitude of the primordial curvature power spectrum $10^9 A_s$ and its spectral index $n_s$.
We fix the optical depth to reionization, $\tau=0.055$, because both DES and CMB lensing measurements are insensitive to its value.
In addition to these parameters, we include the standard treatment of systematic effects for DES data, as described in~\cite{DES:2017myr}, that includes:
bias parameters for the lens galaxy sample; intrinsic alignment parameters; parameters describing photometric redshift uncertainties for both sources and lens galaxies; shear calibration parameters.
All cosmological and nuisance parameters have priors that match the ones used in~\cite{DES:2017myr}. 
As derived parameters, we consider the amplitude of the linear power spectrum on the scale of {8~$h^{-1}$~Mpc}, $\sigma_8$.

Throughout this section, we compute cosmological predictions with CAMB~\cite{Lewis:1999bs}, and we Markov chain Monte-Carlo (MCMC) sample the relevant posteriors with CosmoMC~\cite{Lewis:2002ah} until the Gelman-Rubin convergence statistic~\cite{gelman1992} satisfies $R-1<0.01$, or better.
The statistical analysis of the MCMC chains is performed with GetDist~\cite{Lewis:2019xzd} and Tensiometer~\cite{Raveri:2021wfz} that will contain the methods discussed in this paper in a future code release.

In addition to the non-linear analysis, we perform a linear analysis as presented in \cref{Sec:LinearMethods}.
Since all the posteriors we consider for DES are clearly non-Gaussian, the linear analysis is not expected to work well, but is still instructive. 
In order to get the most out of the linear analysis, and due to our prior knowledge that we will want to relate certain parameters by power-law decompositions, we conduct our linear analysis on the parameters in log space. 
We perform the linear analysis on the local Fisher matrix, as estimated from the normalizing flow model, computed at the mean of the posterior.
This is in contrast to \cref{Sec:LinearMethods} where the Gaussian distributions allowed us to use the global Fisher matrix in our analysis. We have verified that using the local Fisher matrix at the mean is the most stable choice as it helps mitigate the effects of non-Gaussianities on the results.
In fact, when using and comparing the prior and posterior distribution covariances around their means, one implicitly compares covariances at different points in parameter space. This would yield the same results in the Gaussian case but not in the non-Gaussian case, hence our choice to work with local covariances evaluated at the same point.

In our linear CPC analyses of the cosmological data, it is also instructive to calculate the number of relevant data constrained modes.
For $N$ varied parameters, this is given by~\cite{Raveri:2018wln} as:
\begin{align} \label{Eq:NeffDefinition}
N_{\rm eff} \equiv N - {\rm Tr}\left( \mathcal{C}_\Pi^{-1} \mathcal{C}_p\right)
\end{align}
which is the quantity that controls the goodness of fit of the maximum posterior, as shown in~\cite{Raveri:2018wln}.
CPCA studies the posterior with respect to the prior, so this is a useful number to gauge how many modes we should consider.
Note that a similar proxy does not exist for PCA.
Using the CPC decomposition, we can write $N_{\rm eff}$ as a sum of independent terms as:
\begin{align} \label{Eq:NeffSpectrum}
N_{\rm eff} = N - {\rm Tr}(\Lambda^{-1}) = \sum_{i=1}^N 1 - \lambda_i^{-1}
\end{align}
and analyze the spectrum of $N_{\rm eff}$ to understand how many modes are only partially data constrained.

\subsection{DES  cosmic shear and 3x2} \label{Sec:DES.shear}
We start with the analysis of the $\Lambda$CDM posterior of DES Y1 cosmic shear and DES Y1 so-called 3x2 measurements, combining galaxy lensing and galaxy clustering data.
In both cases, all parameters describing photometric redshift uncertainties and shear calibration are completely prior constrained so that we can avoid including them in the analysis and just consider the posterior distribution that is obtained marginalizing over them. 
Hence, for shear, we consider all cosmological parameters and intrinsic alignment parameters, $A_{\rm IA}$ for the amplitude of intrinsic alignment and $\alpha_{\rm IA}$ for its redshift dependence.
In the 3x2 case, we instead consider all cosmological parameters, IA parameters and five bias parameters, one for each galaxy clustering redshift bin.

We can start by computing $N_{\rm eff}$, the number of relevant data constrained modes, following \cref{Eq:NeffSpectrum}.
For shear, this procedure gives ${N_{\rm eff} = \sum (1,\, 0.89,\, 0.58,\, 0.36,\, 0.12,\, 0,\,0) = 2.95}$ and tells us that one mode is very well constrained over the prior, while three other modes are less constrained and the prior is expected to influence them at a level that is greater than $10\%$, and two modes are completely prior-constrained.

\begin{table}
\setlength{\tabcolsep}{3pt}
\centering
\begin{tabular}{ccccccccc}
\multicolumn{9}{l}{DES Y1 shear full parameter space, $N_{\rm eff} = 2.95$} \\
\toprule
Mode & $\sqrt{\lambda_i-1}$ & $\Omega_m$  & $\sigma_8$ & $\Omega_b$ & $H_0$ & $n_s$ & $A_{\rm IA}$ & $\alpha_{\rm IA}$ \\
\colrule                                               
1 & $38.75$ & $0.31$ & $0.69$ & $0$    & $0$ & $0$    & $0$    & $0$ \\
2 & $6.89$  & $0.01$ & $0.02$ & $0$    & $0$ & $0.01$ & $0.94$ & $0.02$ \\
3 & $1.34$  & $0.6 $ & $0.25$ & $0.04$ & $0$ & $0.06$ & $0.03$ & $0.02)$ \\
\botrule
\end{tabular}
\caption{\label{Tab:DESY1.shear.contribution}
DES Y1 shear contribution of parameters to the variance of each CPC mode that is significantly constrained over the prior.
The second column shows the posterior to prior variance improvement for each mode. }
\end{table}

In \cref{Tab:DESY1.shear.contribution}, we show the contribution of each model parameter to the variance of each CPC mode. 
As discussed in \cref{Sec:KL}, the eigenvalues corresponding to CPC modes quantify the improvement of the posterior with respect to the prior. 
Each mode in the table is listed with the value of the posterior to prior variance improvement. 
For the full set of parameters, we can see the first mode involves only $\sigma_8$ and $\Omega_m$, the second mode contains most of the IA contribution with little to no degeneracy with cosmological parameters; while the third mode is again dominated by $(\sigma_8, \Omega_m)$ but is largely influenced by the prior.

Based on these results we can marginalize over IA parameters and consider only cosmological parameters in our main analysis.
Since IA parameters are not correlated with cosmological parameters, this marginalization would not affect the end results.
Once we marginalize, we can obtain $N_{\rm eff} = \sum (1,\, 0.58,\, 0.12,\, 0,\, 0) = 1.7$. This is compatible with the previous result since we have removed one IA constrained mode.
The lack of correlation between IA parameters and cosmological parameters is confirmed by inspection of the shear posterior.

\begin{figure}[h]
\centering
\includegraphics[width=\columnwidth]{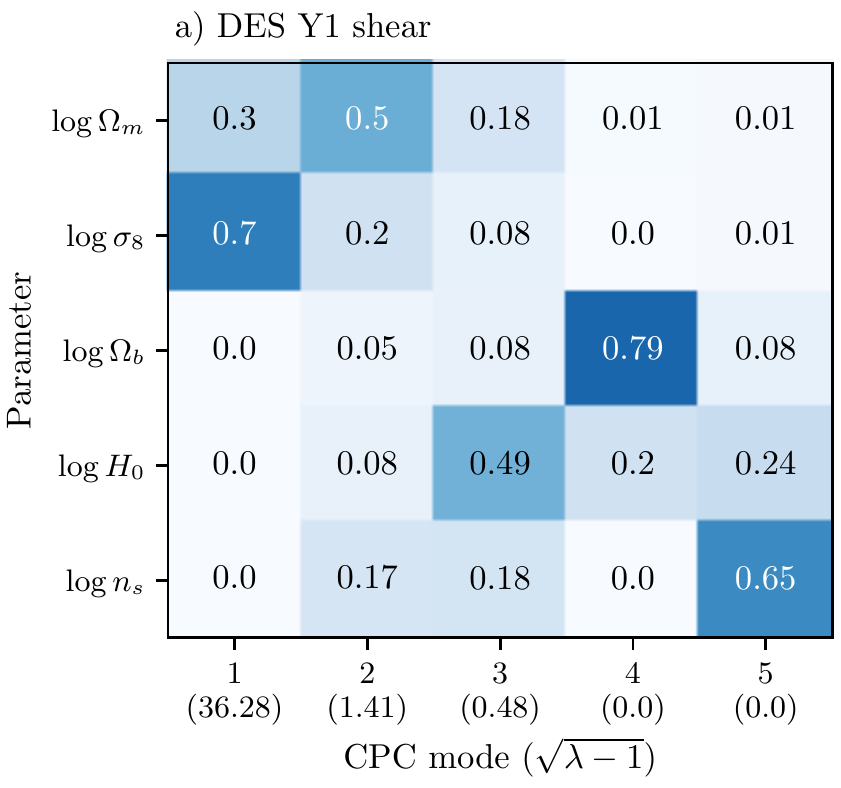}
\includegraphics[width=\columnwidth]{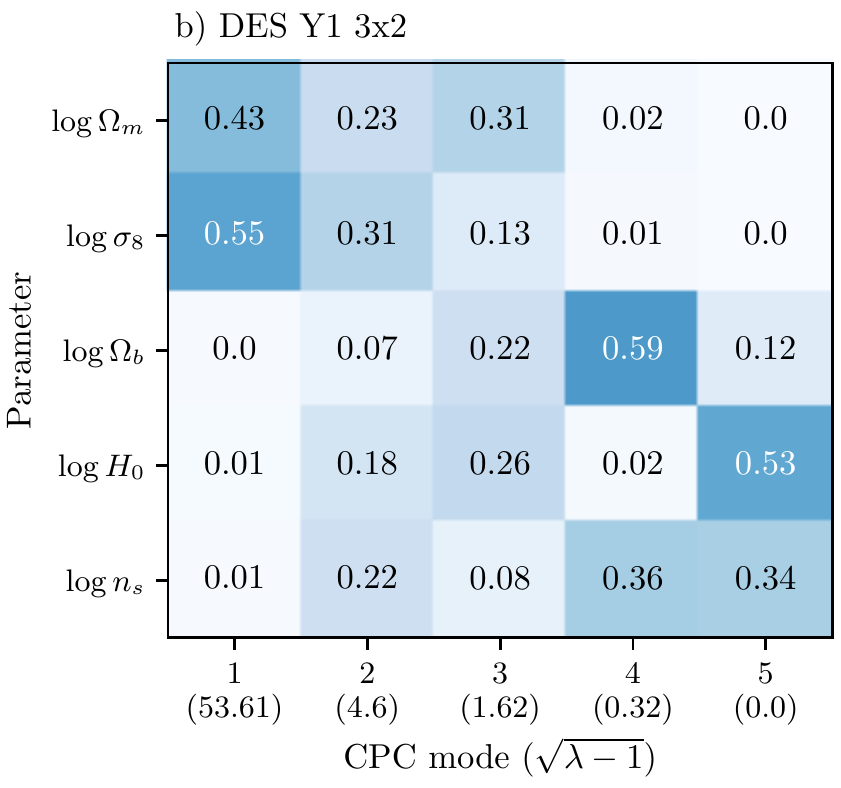}
\caption{ \label{Fig:CPCA.contributions}
Fractional contribution of each cosmological parameter to the variance of each CPC mode for DES Y1 shear (upper) and 3x2 (lower) measurements.
The value in parenthesis on the x-axis indicates the improvement of the posterior with respect to the prior for each mode.
}
\end{figure}
In Panel~a) of \cref{Fig:CPCA.contributions}, we show the posterior to prior improvements and the contributions of each cosmological parameter to the variance of each CPC mode. 
As we can see, for the first mode, the improvement of the posterior with respect to the prior is significant.
The composition of both the first and second modes do not appreciably change with respect to the first and third modes in \cref{Tab:DESY1.shear.contribution}. This is a consequence of the lack of correlation between IA and cosmological parameters. The first and second mode are both dominated by $\sigma_8$ and $\Omega_m$, but the second mode also includes non-negligible contributions from $H_0$ and $n_s$. Note that this CPC analysis correctly classifies the $n_s$ parameter as dominant only in the least constrained mode. 

We can similarly consider the $\Lambda$CDM posterior of the DES Y1 3x2 measurement.
In this case, for the full set of parameters, $N_{\rm eff} = \sum (1,\, 1,\, 1,\, 1,\, 0.99,\, 0.99,\, 0.98,\, 0.9,\, 0.38,\, 0.32,\, 0.1,\, 0) = 8.64$. This tells us that four modes are extremely well constrained over the prior, four modes are very well constrained over the prior, three modes are much less constrained and the prior influences them significantly, and one mode is completely prior constrained. 

Similarly to the previous case, we find that combinations of bias parameters are uncorrelated with cosmological parameters of interest.
This is due to the presence of the galaxy-lensing cross-correlation in the 3x2 data which breaks the $\sigma_8$-bias degeneracy. 
Marginalizing over all parameters besides cosmological parameters in DES Y1 3x2, we get the following decomposition of ${N_{\rm eff} = \sum (1,\, 0.95,\, 0.49,\, 0.1,\, 0) = 2.54}$. This indicates that there are two relevant modes that are not significantly influenced by the prior. This is one more constrained mode than in of DES Y1 shear. We examine the improvement over the prior of each mode as well as the breakdown of parameter contributions for 3x2 in Panel~b) of \cref{Fig:CPCA.contributions}. 

\begin{figure*}[hp!]
    \centering
    \includegraphics[width=.95\columnwidth]{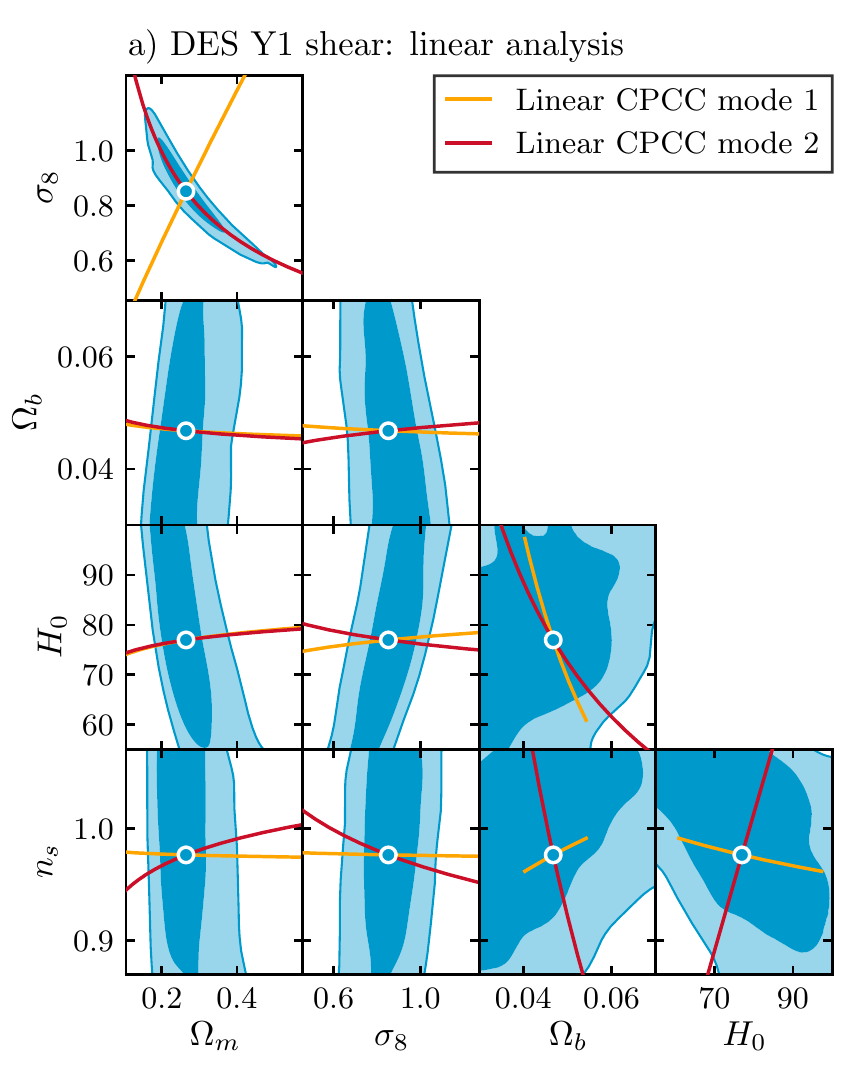}
    \includegraphics[width=.95\columnwidth]{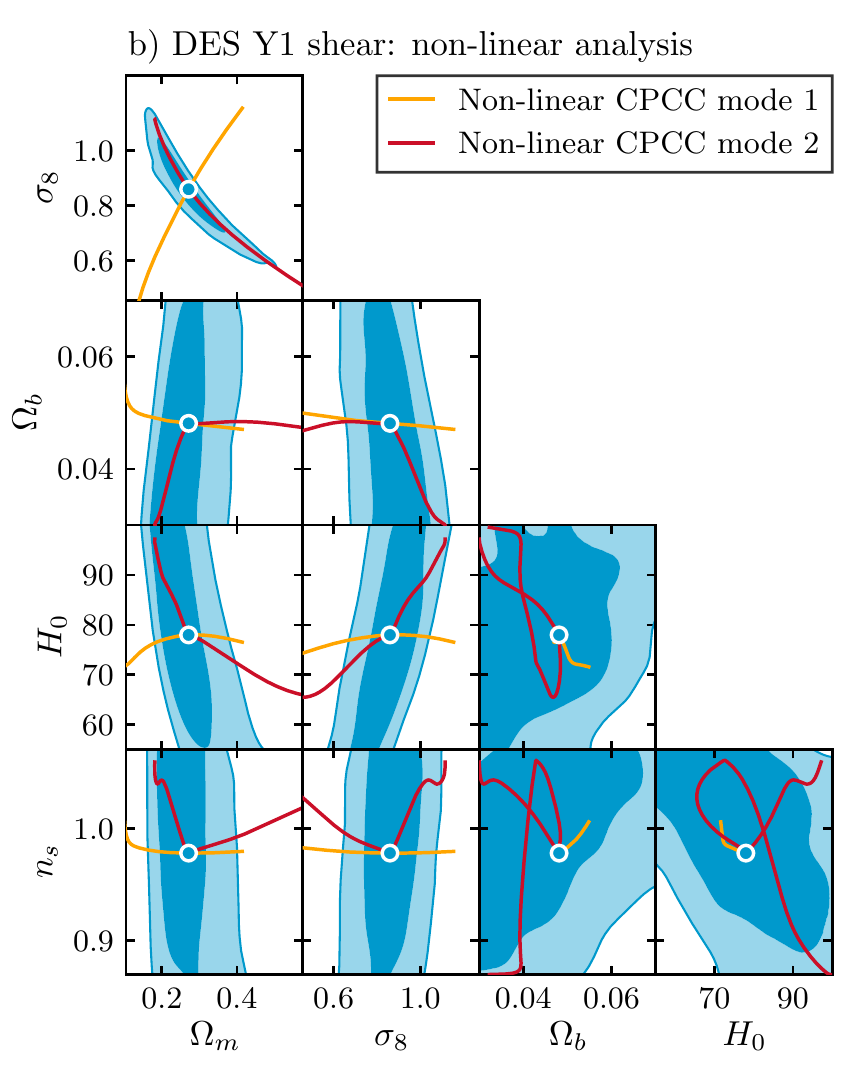}
    \includegraphics[width=.95\columnwidth]{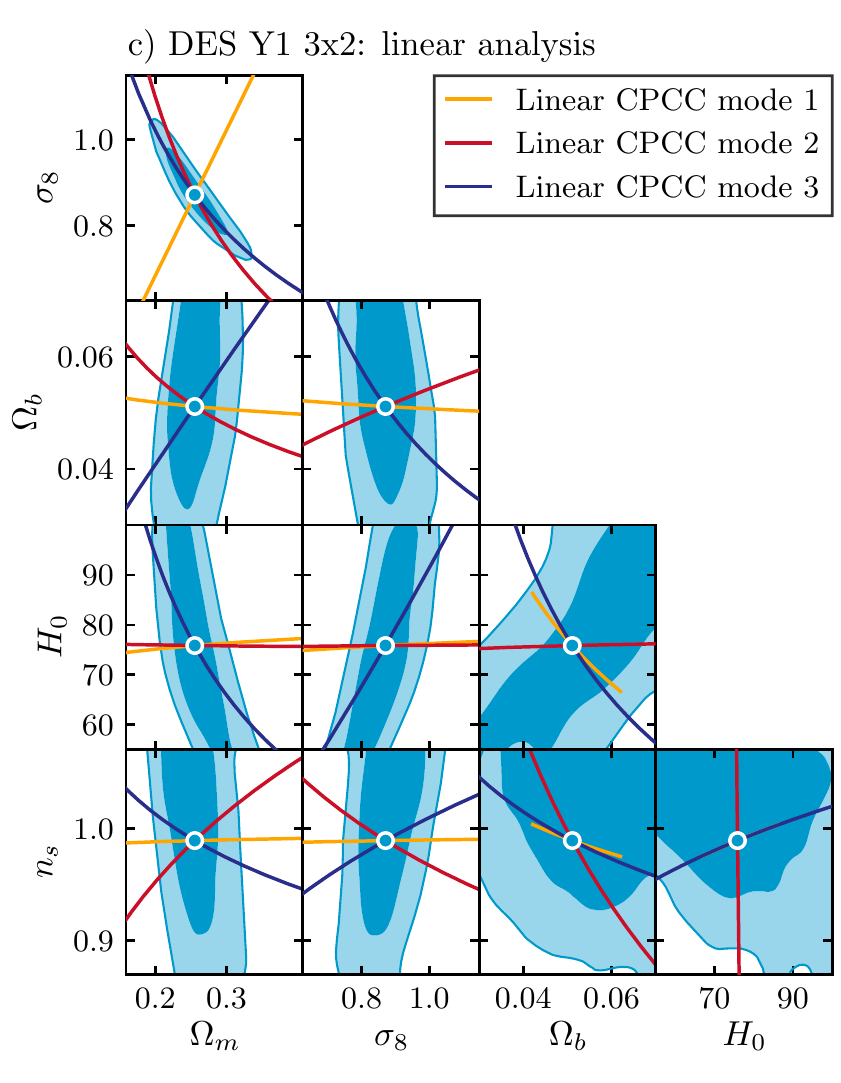}
    \includegraphics[width=.95\columnwidth]{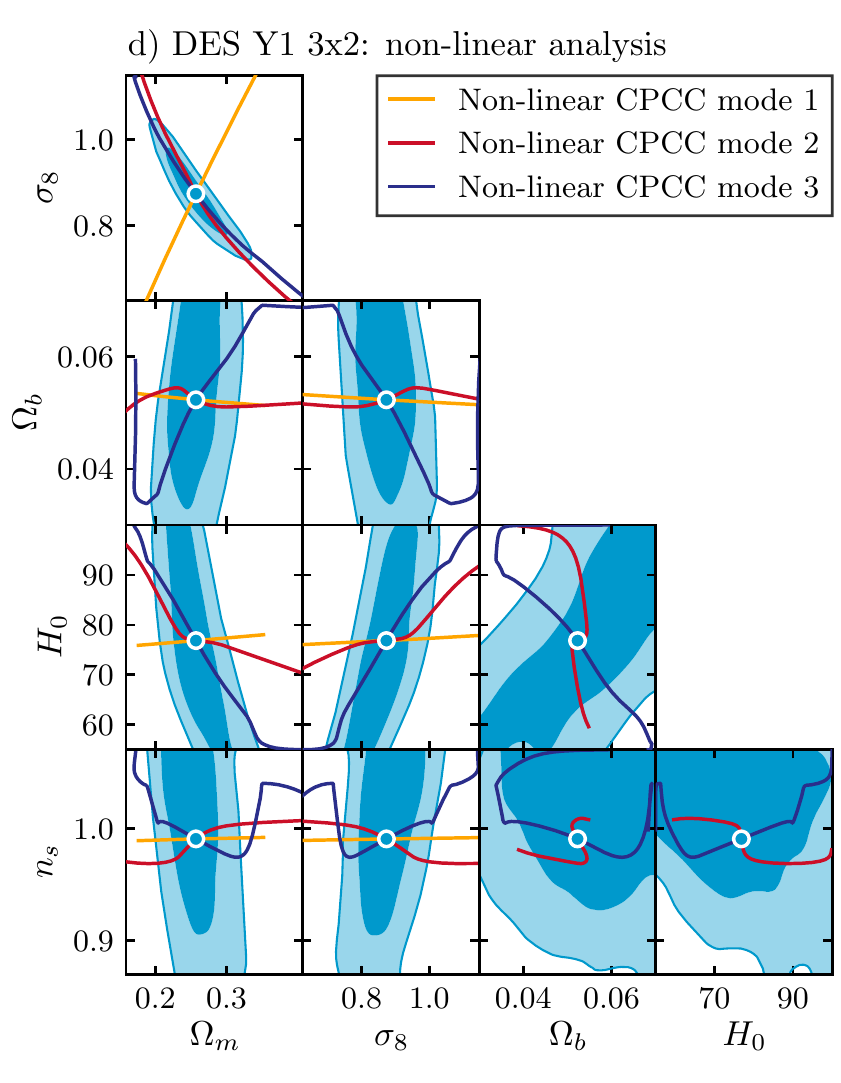}
    \caption{ \label{Fig:LCDM.triangles}
    Linear and non-linear CPCA curves for DES Y1 shear (upper) and 3x2 (lower) measurements.
    In all panels different colors correspond to different modes, as shown in legend.
    The darker and lighter shades correspond to the 65\% and 98\% C.L. regions.
    }
\end{figure*}
The first mode of DES Y1 3x2 shows more improvement over the prior than the first mode of DES Y1 shear. 
For the second mode we also see, as we would expect from the $N_{\rm eff}$ results, that the 3x2 data combination constrains it more significantly than in shear. 
As seen with DES Y1 shear, the first mode is almost entirely dominated by contributions from $\sigma_8$ and $\Omega_m$. 
The second relevant mode for 3x2, however, involves many parameters and in particular has some contribution from $n_s$ and $H_0$, indicating that its constraint might be influenced by the prior on those two parameters.
This influence is expected to be sub-leading though, since the spectral $N_{\rm eff}$ analysis indicates that the prior contributes at most 5\% to this mode.
Once again, the CPC analysis correctly classifies $n_s$ as contributing mostly to the two least constrained modes.

In \cref{Fig:LCDM.triangles}, we show the linear and non-linear CPCC modes for the DES Y1 shear and 3x2 measurements. We plot one more mode with respect to what is effectively constrained by data, as determined by the value of $N_{\rm eff}$. 
For the shear linear analysis in Panel a) of \cref{Fig:LCDM.triangles}, the first mode visibly captures the most constrained direction in the $\sigma_8$ and $\Omega_m$ space, and is mostly constant when projected on the other parameters. The second mode traces the degeneracy in the $\sigma_8$ and $\Omega_m$ space and, while mostly constant in other parameters, shows the increased contribution of $n_s$ and $H_0$. 
The non-linear analysis for shear in Panel b) of \cref{Fig:LCDM.triangles} results in similar modes to the linear analysis for the most constrained parameters, tracing those distributions as expected. 
Note that the apparent coiled shape and short length of the Covariant Principal Component Curves in certain parameters is due to the projection of a curve in five-dimensional space in two dimensions.
We can also note a highly non-linear structure that results in marked differences with the linear analysis for parameters that contribute very little to a given mode and are prior constrained, as, for example, in the $H_0, \Omega_b$ panel.

In the lower panels of \cref{Fig:LCDM.triangles}, we show the linear and non-linear CPCC modes for the DES Y1 3x2 measurements. For the linear analysis, the first mode still visibly captures the most constrained direction in the $\sigma_8$ and $\Omega_m$ space and is still mostly constant when projected on other parameters. The second mode still traces the degeneracy in the $\sigma_8$ and $\Omega_m$ space and shows the contributions of $n_s$ and $H_0$. The third mode still captures the degeneracy in $\sigma_8$ and $\Omega_m$ but also folds in the degenerate directions of $H_0$ and $\Omega_b$. 
The non-linear analysis in Panel d) of \cref{Fig:LCDM.triangles} recovers similar curves to the linear analysis, particularly in the inner regions.

We can then use the linear analysis to gain intuition on the structure of the best constrained modes. 
That is, we can use \cref{Eq:PlaneKL} to define a hyper-plane orthogonal to a given constrained CPC mode. 
We consider these hyper-planes to be an analytical description of a given mode. 
In our analytical expressions, we ignore the parameters that do not contribute over 10\% to the variance of the mode.
The only constrained mode for shear can therefore be described by:
\begin{align}
p_1 = \sigma_8 \Omega_{m}^{0.56} \ \ \ {\rm (CPCA: Cosmic\ Shear)} \,,
\end{align}
This exponent only changes very slightly with and without marginalization over IA parameters and agrees with the standard definition of $S_8 \equiv \sigma_8 \Omega_m^{0.5}$~\cite{Jain:1996st}.

\begin{figure}
\centering
\includegraphics[width= \columnwidth]{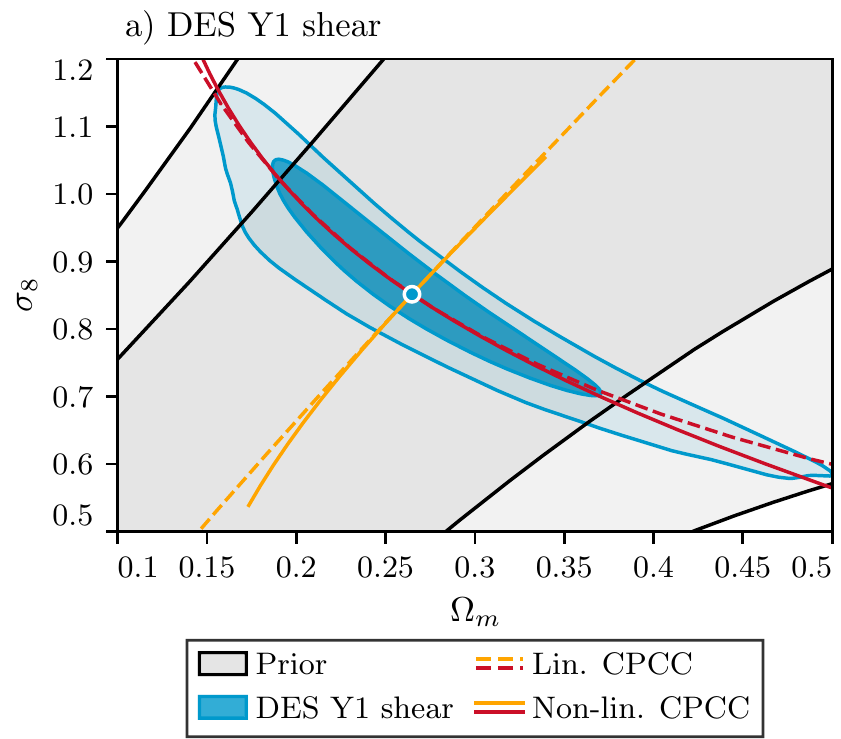}
\includegraphics[width= \columnwidth]{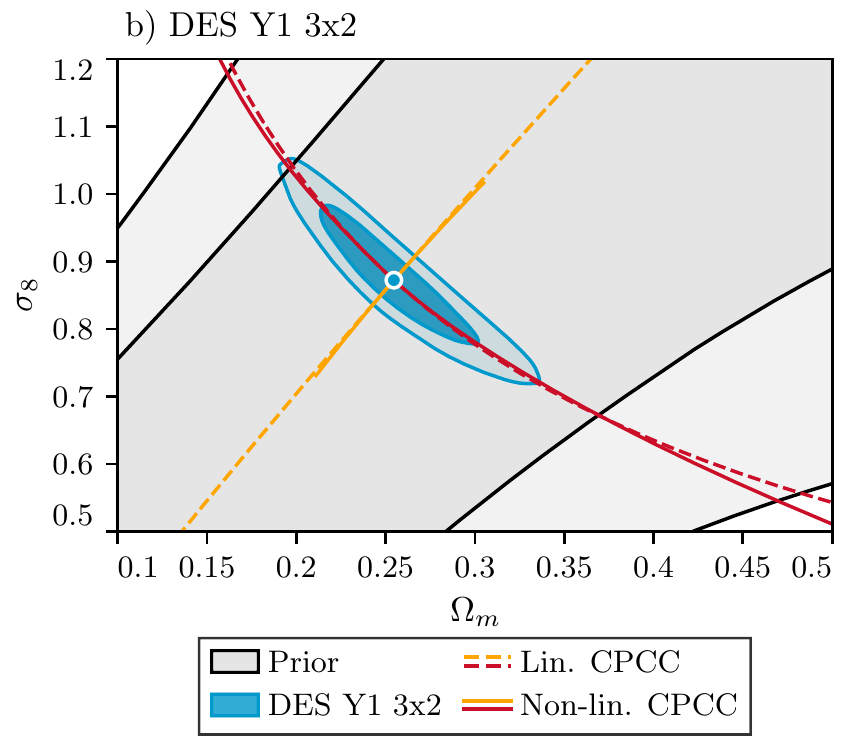}
\caption{ \label{Fig:LCDM.single}
The 2d joint marginalized posterior of $\sigma_8$ and $\Omega_m$ for DES Y1 shear (left) and 3x2 (right). 
We show linear and non-linear Covariant Principal Component Curves (CPCC) calculated from the mean of the distribution with dashed and continuous lines respectively.
The darker and lighter shades correspond to the 65\% and 98\% C.L. regions.
}
\end{figure}

For the two constrained modes in DES Y1 3x2, we obtain the following expressions:
\begin{align}
p_1 =& \sigma_8 \Omega_{m}^{0.89} \ \ \ {\rm (CPCA: 3x2)}\,, \nonumber \\
p_2 =& \sigma_8 \Omega_{m}^{-0.91} n_s^{-1.3} H_0^{-1.0} \,.
\end{align}
Unlike with DES Y1 shear, analytical expression describing the most constrained mode does not recover the standard definition of $S_8$. The hyper-plane involves a higher power of $\Omega_m$, corresponding to a rotation of the posterior distribution. As $S_8$ does not accurately describe the best constrained mode in DES Y1 3x2, this linear analysis with Covariant Principal Components allows us to recover the correct power-law decomposition that decorrelates the parameters $\sigma_8$ and $\Omega_m$. 
The analytical expression describing the second mode is significantly more complicated as it mixes in contributions from many parameters.

Since the first modes of both DES Y1 shear and 3x2 do not depend on other cosmological parameters besides $(\sigma_8, \Omega_m)$, we can marginalize over all other parameters and consider only these two. 
In \cref{Fig:LCDM.single}, we show the linear and non-linear CPC analysis over-plotted for the marginalized posterior of $\sigma_8$ and $\Omega_m$ for DES Y1 shear and 3x2. In both cases, the linear and non-linear modes agree exceptionally well, apart from in the most exterior regions. As expected, the first modes define the most constrained direction of the posterior with respect to the prior and the second modes trace the degenerate direction of $\sigma_8$ and $\Omega_m$. 
We can also see that, for the 3x2 analysis, the agreement between the linear decomposition and the fully non-linear one improves, with respect to shear. This is a consequence of the central limit theorem. The 3x2 data combination is more constraining and more Gaussian distributed data modes are projected over the same parameters, so we expect the corresponding posterior to be more Gaussian.

Note that the agreement of the linear and non-linear analysis for our cosmological examples hinges on the fact that we conducted our linear analysis in log space due to prior knowledge of the power-law decompositions we intended to recover and the transformation that would decrease the non-gaussianity of $\sigma_8$ and $\Omega_m$ posterior. 
In general, for non-gaussian distributions, the non-linear analysis would be able to recover the correct curves that capture information about the distribution while the linear analysis would only be a good approximation close to the mean.

\subsubsection{PCA with DES Y1 cosmic shear}
\begin{figure}
\centering
\includegraphics[width=\columnwidth]{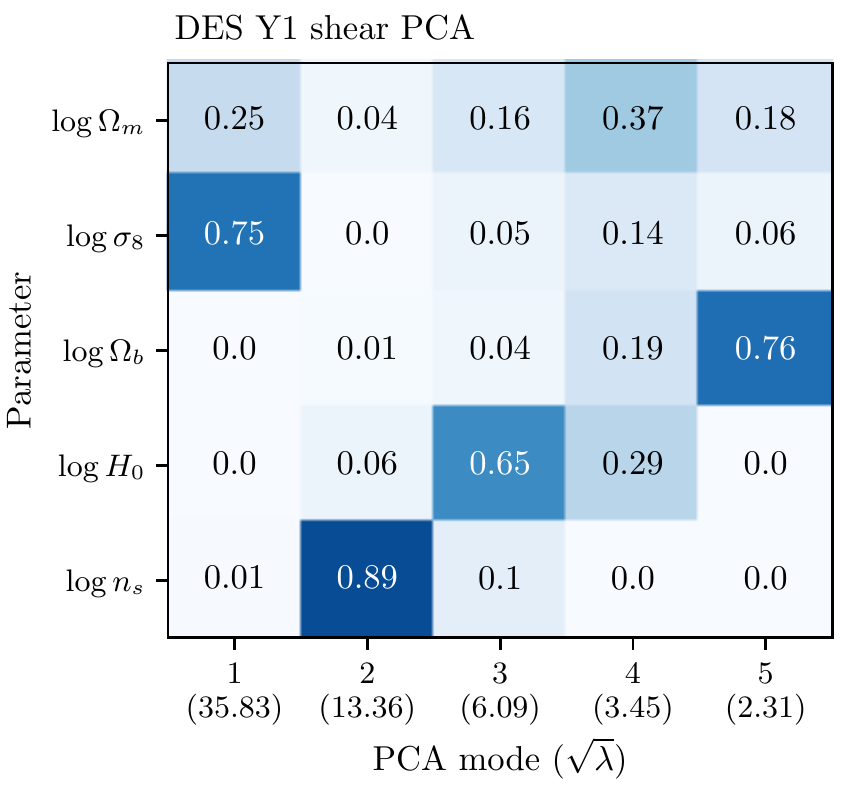}
\caption{ \label{Fig:LCDM.PCA.contributions}
Fractional contribution of each cosmological parameter to the variance of PC modes for DES Y1 shear measurements.
}
\includegraphics[width=\columnwidth]{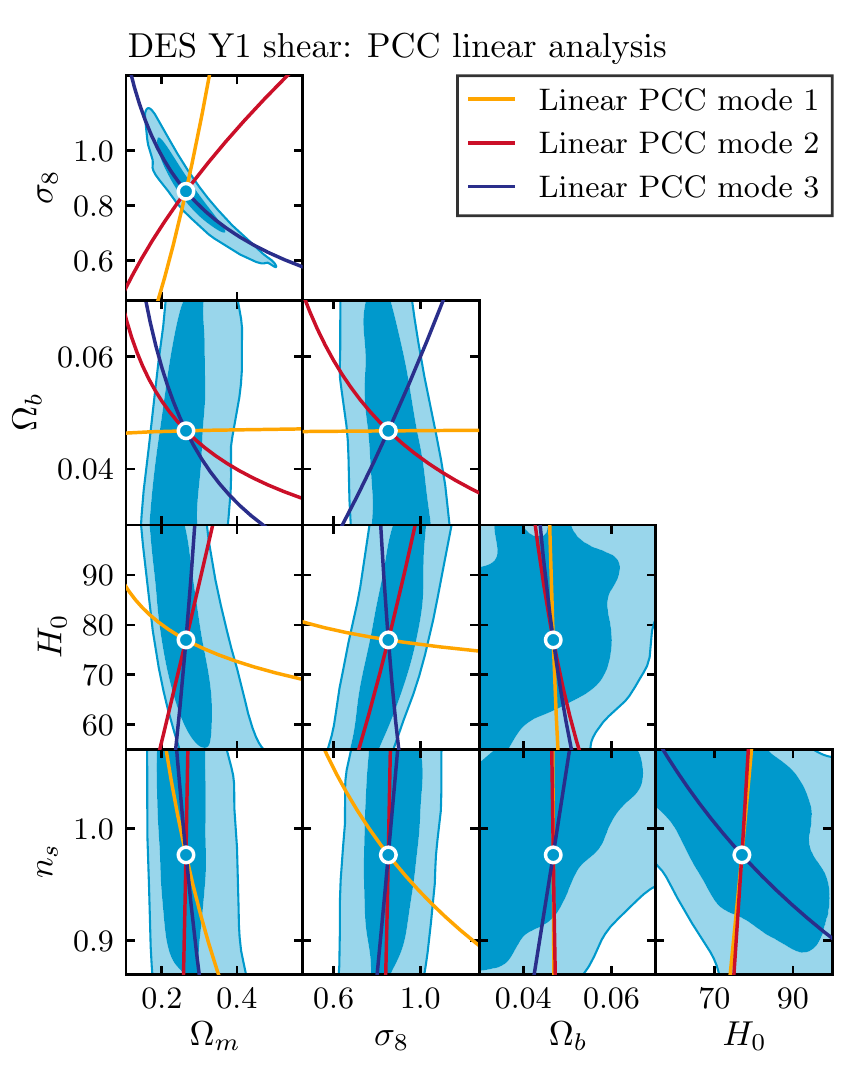}
\caption{ \label{Fig:LCDM.PCA.shear.triangles}
Linear PC modes for DES Y1 shear measurements.
Different colors correspond to different modes, as shown in legend.
The darker and lighter shades correspond to the 65\% and 98\% C.L. regions.
}
\end{figure}

We can now gauge the differences between CPCA and PCA for the DES Y1 shear example in particular.
Considering only cosmological parameters, we can compute the contribution of each parameter to the variance of each PC mode as we did with CPC modes, as shown in \cref{Fig:LCDM.PCA.contributions}.
As we can see, the first mode is still identified as being contributed mostly by $\sigma_8$ and $\Omega_m$.
However, precisely as it happens in the second example discussed in \cref{Sec:ToyExample.PriorInformed}, the second mode is incorrectly identified as being dominated by $n_s$.
This is due to the fact that the prior on $n_s$ is very tight and is misinterpreted by PCA as a data constraint.

We have verified that a similar problem would also happen if we were to include the nusiance parameters, that are tightly prior constrained, in the PCA analysis, instead of marginalizing over them. 
In that case, those parameters would be mixed in the modes and results would make little sense.

In \cref{Fig:LCDM.PCA.shear.triangles} we show the joint parameter posterior for all five cosmological parameters of interest, together with lines corresponding to the first few PC modes. We arbitrarily choose to show three modes, since there is no quantity corresponding to $N_{\rm eff}$ that can be calculated for PCA. 

The first mode is in qualitative agreement with the first CPC mode, shown in the previous section.
This result, however, should be regarded as a coincidence due to the choice of parameters that we are using.
The second mode, while it appears to capture the most constrained direction in $\sigma_8$ and $\Omega_m$ space, is actually the mode dominated almost entirely by $n_s$ and has little contribution from $\Omega_m$ and no contribution from $\sigma_8$.

\subsection{DES Y1: from cosmic shear to 3x2} \label{Sec:DES.shear.to.3x2}
If we have two data sets that are nested, i.e. one is part of the other, we can use the same methods we have discussed to understand what is improved in the joint data combination.
To do so, we simply treat the first experiment as the prior for the joint data set.
In this section, we show a worked example based on DES Y1 measurements, considering shear as a prior and the 3x2 posterior as the posterior.
The DES 3x2 data vector in fact contains shear measurements and adds galaxy clustering and its cross-correlation with shear measurements.
In this case, we need to consider the full space of all parameters that the two experiments share.

As in the previous section we can compute ${N_{\rm eff} = \sum (0.97,\, 0.9,\, 0.4,\, 0.15,\, 0.07,\, 0,\, 0) = 2.49}$ over the full cosmological and IA parameter space.  This is telling us that two modes are significantly improved by 3x2 over shear.

\begin{figure}
\centering
\includegraphics[width=\columnwidth]{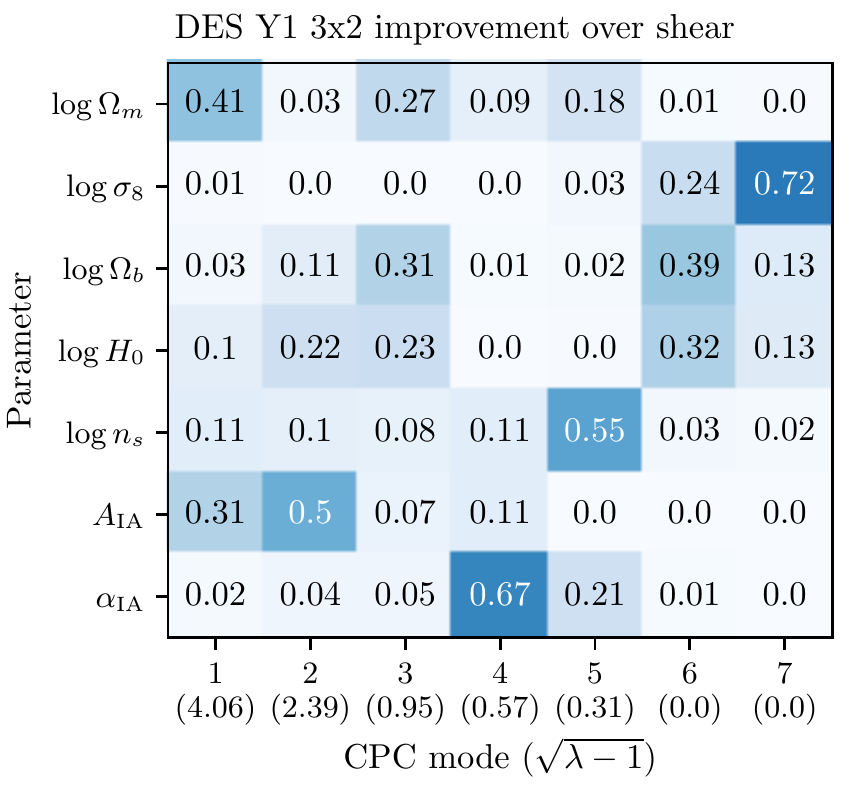}
\caption{ \label{Fig:DESY1.shear.to.3x2.contribution}
Contribution of parameters to the variance of each CPC mode that the DES Y1 3x2 measurement improve over shear measurements.
The number in parenthesis on the x-axis shows the 3x2 to shear variance improvement for each mode.
}
\end{figure}

In \cref{Fig:DESY1.shear.to.3x2.contribution} we show which parameters are most improved when going from shear measurements to 3x2.
As we can see, most of the improvement comes from a better constraint on $\Omega_m$ and $A_{\rm IA}$ for the first mode, while the second mode picks up improvement in $A_{\rm IA}$, $H_0$, $\Omega_b$ and $n_s$. Since DES Y1 3x2 builds off of DES Y1 shear by adding galaxy clustering and galaxy-lensing correlations (2x2), and since we marginalize over all bias parameters, we can see that there is no added constraint on $\sigma_8$ when moving from shear to 3x2 due to 2x2.
Also note that we can read off \cref{Fig:DESY1.shear.to.3x2.contribution} the quantitative improvement corresponding to each mode that shows that the variance of the first mode is improved by a factor 4 while the second is improved by a factor 2.
This quantification of improvement is parameter invariant.

\begin{figure}
\centering
\includegraphics[width=\columnwidth]{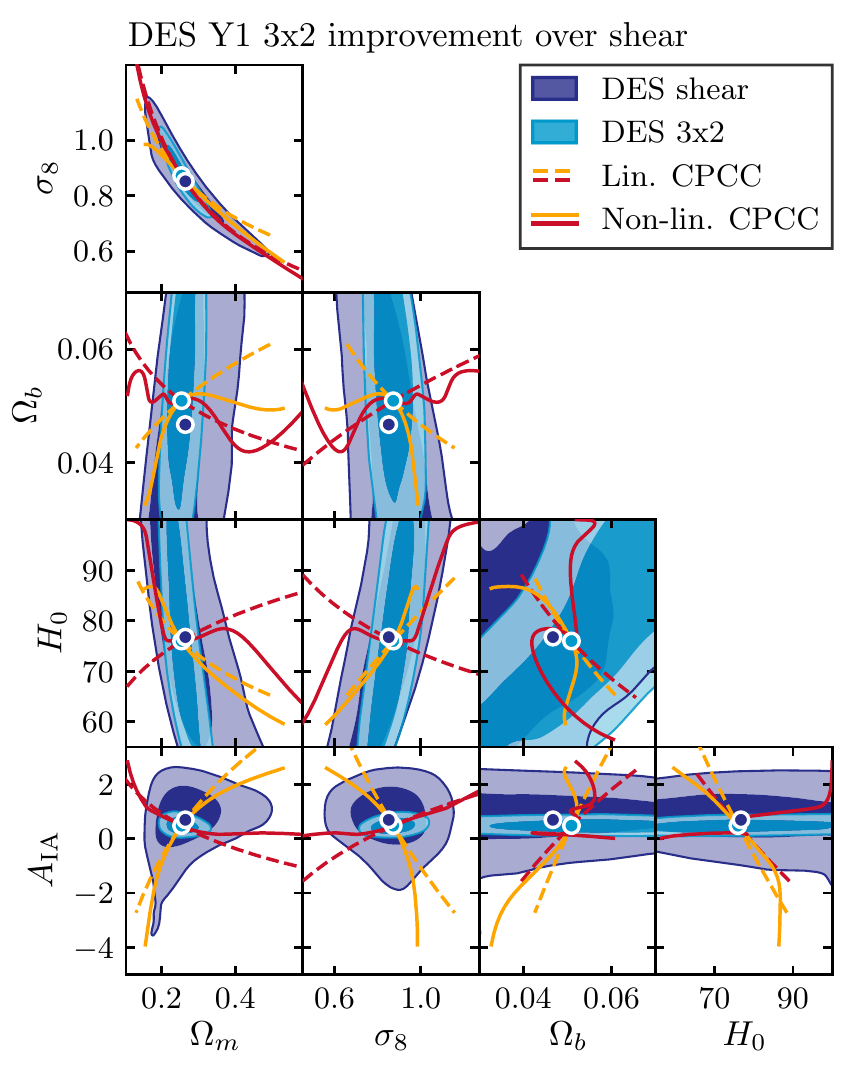}
\caption{ \label{Fig:LCDM.CPC.shear.to.3x2.triangle}
Linear and non-linear CPC modes for the improvement of DES Y1 3x2 measurements over shear measurements.
Different colors correspond to different modes, as shown in legend.
The darker and lighter shades correspond to the 65\% and 98\% C.L. regions.
}
\end{figure}

We examine the actual directions of improvement found using our linear and non-linear analysis in \cref{Fig:LCDM.CPC.shear.to.3x2.triangle}. 
In this figure, we discard parameters that are not relevant to our discussion. We note that the improvement in $n_s$ from the second mode appears sub-dominant and is likely classified as above $10\%$ in \cref{Fig:DESY1.shear.to.3x2.contribution} because of some noise in the posterior distributions. We therefore do not include $n_s$ in \cref{Fig:LCDM.CPC.shear.to.3x2.triangle}.  
In this case, we can notice that the non-linear structure of the modes is significant. 
It is clearly shown by the $A_{\rm IA}$, $\Omega_m$ panel of \cref{Fig:LCDM.CPC.shear.to.3x2.triangle}. Around the mean, the two analyses agree, while they rapidly differ when moving away from the mean.
As in the previous plots parameters that do not project significantly on the constrained modes are noisy.

\subsection{DES Y1 and CMB lensing} \label{Sec:DES.CMB.lensing}
In this section we show how to use the CPCA decomposition to understand fruitful synergies between different experiments.
To do so, we consider CMB lensing reconstruction measurements from Planck~\cite{Planck:2018lbu}.
In this case, we can consider only cosmological parameters since the only nuisance parameter, describing the overall calibration, is completely prior constrained.
The number of effective parameters that are constrained is given by $N_{\rm eff} = \sum (0.99,\, 0.96,\, 0.56,\, 0,\, 0) = 2.51$, showing that we have two constrained modes.

\begin{figure}
\centering
\includegraphics[width=\columnwidth]{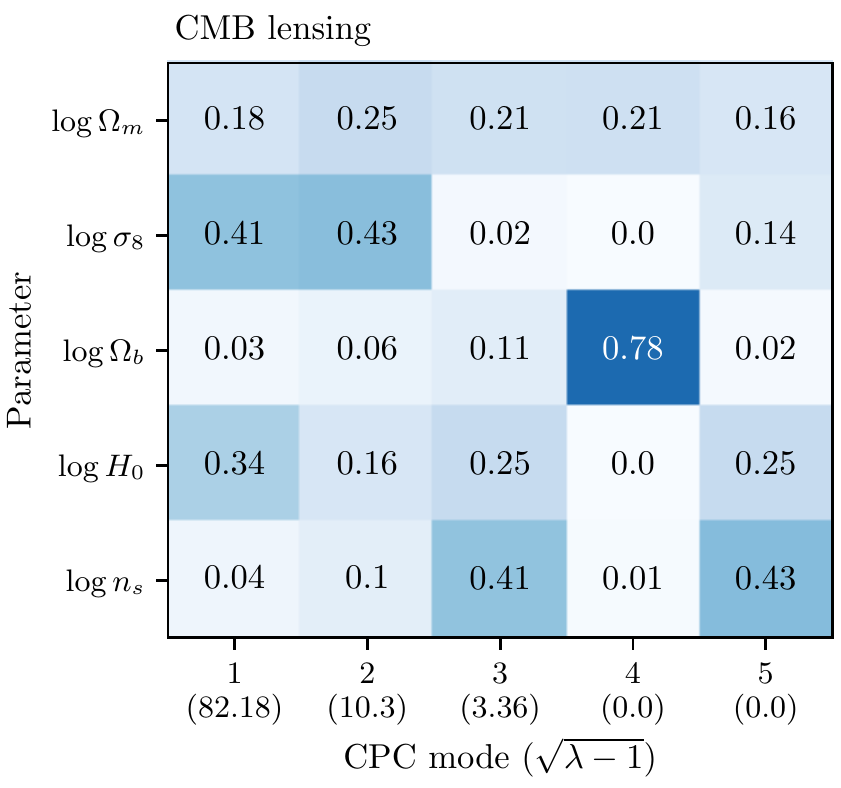}
\caption{ \label{Fig:CMB.Lensing.contributions}
Contribution of parameters to the variance of each CPC mode that is measured by Planck CMB lensing reconstruction.
The number in parenthesis on the x-axis shows the 3x2 to shear variance improvement for each mode.
}
\end{figure}

As we can see from \cref{Fig:CMB.Lensing.contributions} the first two best constrained modes receive most of their contribution from $\Omega_m$, $H_0$ and $\sigma_8$ and a sub-leading contribution from $\Omega_b$ and $n_s$.
This is expected based on the discussion in~\cite{Pan:2014xua, Baxter:2020qlr}.
The amplitude of the CMB lensing signal responds to the overall amplitude of matter perturbations and the angular scale of matter radiation equality $\ell_{\rm eq} \equiv k_{\rm eq} \chi_* \propto \Omega_m^{\alpha} h$, that is fixed to 
$\ell_{\rm eq} \propto \Omega_m^{0.6} h$ in~\cite{Planck:2015mym} with a PCA decomposition.

We can now look at the analytic linear approximation for the modes, finding:
\begin{align} \label{Eq:CMB.Lensing.Mode1}
p_1 =& \sigma_8 \Omega_m^{-0.8} H_0^{-1.0} \ \ \ {\rm (CPCA: CMB\ Lensing)} \,, \nonumber \\
p_2 =& \sigma_8 \Omega_m^{-1.8} H_0^{-2.8}
\end{align}
which has a clear contribution that looks like $\ell_{\rm eq} \propto \Omega_m^{0.8} h$, with a slight different exponent.
Noticeably the best constrained direction is not $\sigma_8 \Omega_m^{0.25}$, as reported in~\cite{Planck:2015mym}, 
and that is due to the addition of an informative prior on $\Omega_b h^2$ in~\cite{Planck:2015mym}.

\begin{figure}
\centering
\includegraphics[width=\columnwidth]{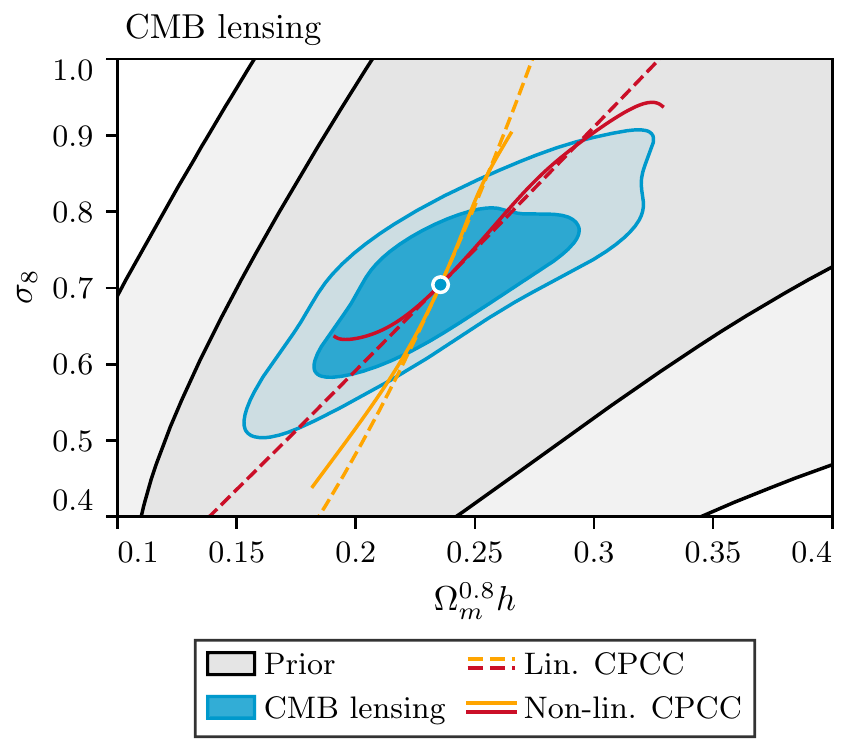}
\caption{ \label{Fig:CMB.Lensing.triangle}
The 2d joint marginalized posterior of $\sigma_8$ and $\Omega_m^0.8 h$ for CMB lensing.
We show linear and non-linear Covariant Principal Component Curves (CPCC) calculated from the mean of the distribution with dashed and continuous lines respectively.
The darker and lighter shades correspond to the 65\% and 98\% C.L. regions.
}
\end{figure}
In \cref{Fig:CMB.Lensing.triangle} we show the 2D joint posterior of the parameters involved in CMB lensing that we identified most contribute to the constrained modes.
We can observe that the prior is relevant in the direction that we would associate with the best constrained direction by eye.
On the other hand, the true best constrained mode tracks the direction that is most improved over the prior. As we can also see non-linearities in the first mode are limited, as we would expect for a well constrained mode.
The second mode shows a clear non-linearity. This is due to the fact that we have projected out $n_s$ that has a sub-leading effect that becomes relevant in the tails of the distribution.

Once we understand what CMB lensing is measuring we can pick data combinations that would yield results of interest.
As an example, if we are interested in the value of the Hubble constant, we see that we can extract it from the first CMB lensing mode, once we break its degeneracy with $\Omega_m$ and $\sigma_8$.

The authors of~\cite{Baxter:2020qlr} decide to break the degeneracy using a CMB inspired prior on $A_s$ (and hence a prior on $\sigma_8$) and SN measurements from Pantheon~\cite{Pan-STARRS1:2017jku} for $\Omega_m$ to obtain $H_0 = 73.5 \pm 5.3$ km/s/Mpc.
We can alternatively use redshift space distortion measurements~\cite{Gil-Marin:2015sqa} to constrain $\sigma_8$, 
this would break the degeneracy with $\Omega_m$ for DES 3x2 measurements, that we discussed in the previous section, giving $H_0 = 73.8 \pm 7.5$ km/s/Mpc once joined with CMB lensing.
If we were to further add Pantheon SN we would get $H_0 = 73.0 \pm 6.6$ km/s/Mpc.

This measurement of the Hubble constant is given by the angular size of the matter-radiation equality peak in the power spectrum that is calibrated by radiation density from the measurement of the CMB temperature monopole.
Once the leading degeneracies are broken we are left with residual degeneracies, in particular between $\Omega_b h^2$ and $h$. Once we break this degeneracy further we can improve on the measurement of the Hubble constant, using $\Omega_b h^2$ measurements from BBN~\cite{Cooke:2016rky}, as the absolute calibrator, as it is done in~\cite{DES:2017txv}.

\section{Conclusions} \label{Sec:Conclusions}
As the complexity of cosmological data analysis inevitably increases with more ambitious surveys, so does the need to develop methods to disentangle this complexity and understand what aspects of a given model are measured by data.
In this paper, we have developed such methods, and applied them to examples from current cosmological surveys.

After reviewing the standard procedure of Principal Component Analysis (PCA) and highlighting its problems with lack of covariance under changes of parameter bases, we showed how to perform a decomposition of parameter space that
is covariant under parameter changes.
We have called this decomposition Covariant Principal Component Analysis (CPCA).
By comparing the covariance of the prior distribution to the covariance of the posterior distribution, this decomposition allows us to extract the parameter combinations that the data  improves over the prior.
These parameter combinations, or modes, are statistically independent and are sorted by data improvement over the prior, in a classification that is parameter invariant.
The modes that are well-measured over the prior are also the least influenced by projection effects in high dimensional parameter spaces, a useful feature since these effects increasingly complicate the interpretation of parameter posteriors.

We have studied the linear CPCA problem, which applies to Gaussian distributions, discussed its properties and then  generalized the approach to non-Gaussian distributions.
To do so, we have exploited machine learning models, normalizing flows,  for the probability distributions involved.
That allowed us to obtain the local covariance matrix of the distributions at different positions in parameter space and use its inverse, the Fisher matrix, as a local metric in parameter space.
We have shown how to phrase the treatment of non-linearity in the CPC analysis as a problem of transport in non-inertial reference frames.

We have applied both PCA and CPCA to a set of benchmark examples of non-Gaussian distributions, in \cref{fig:toy_example_3} and \cref{fig:toy_example_4},
to highlight the limits in which they would agree, and the cases in which a failure mode of PCA is remedied by CPCA.

We  apply these techniques to cosmological examples from the first year data release of the Dark Energy Survey (DES) and CMB lensing reconstruction from Planck.
We have considered DES measurements of galaxy lensing shear 2-point correlations, cosmic shear, as well as the set of three shear and galaxy clustering 2-point correlations, called the 3x2 datavector.
This allowed us to show several relevant results:
\begin{itemize}
\item Cosmic shear  constrains one parameter combination over the prior while 3x2  constrains two parameter combinations that are relevant for cosmology.
\item The most constrained mode for cosmic shear and 3x2 is different: while it closely agrees with the definition of $S_8\equiv \sigma_8 \Omega_m^{0.5}$ for shear, it does not for 3x2, for which the best constrained mode is $S_8 \equiv \sigma_8 \Omega_m^{0.89}$.
\item The second constrained mode for 3x2 is a non-trivial combination of multiple, degenerate cosmological parameters, highlighting the need for full-parameter space methods.
\item The non-linear structure of the best constrained modes for both shear and 3x2 agrees well with the linear analysis in the bulk of the posterior and only shows deviations from linearity in the tails of both distributions.
\item 3x2 measurements improve two parameter combinations over cosmic shear -- these mostly involve the amplitude of the intrinsic alignment signal and $\Omega_m$.
\item CMB lensing measurements best constrain the parameter combination $\sigma_8 \Omega_m^{-0.8} H_0^{-1}$ that captures the degeneracy between the amplitude of matter perturbations and the angular location of the matter-radiation equality peak of the power spectrum.
\item Once the CMB lensing degeneracy is identified we can select data combinations that break it to give a measurement of the Hubble constant. We combine CMB lensing, 3x2 DES data and RSD measurements from BOSS, yielding $H_0 = 73.8 \pm 7.5$ km/s/Mpc in agreement with \cite{Baxter:2020qlr}.
\end{itemize}

Looking forward, several improvements to the techniques we discuss would be worth pursuing.
The transport problems involved in the non-linear analysis are challenging in practice and progress in this direction would include the development of stable and efficient algorithms for their numerical solution.
Once non-linear modes are obtained they are hard to interpret due to the lack of a simple criterion to write analytic approximations that are available for the linear analysis.
Future work could add to the non-linear analysis symbolic regression algorithms to obtain analytic approximations that would give further insight on the parameters that most contribute to these modes.

\begin{acknowledgments}

We thank 
Gary Bernstein, Mike Jarvis and Shivam Pandey 
for helpful discussions.

MR is supported in part by NASA ATP Grant No. NNH17ZDA001N, and by funds provided by the Center for Particle Cosmology.

Computing resources were provided by the University of Chicago Research Computing Center through the Kavli Institute for Cosmological Physics at the University of Chicago.

\end{acknowledgments}

\appendix

\section{PCA and SNR} \label{Sec:PCA.SNR}
In this appendix we comment on the relation between PCA and the signal-to-noise ratio (SNR) of the posterior parameters with respect to a fixed reference $\theta_0$:
\begin{align} \label{Eq:SNR}
{\rm SNR}^2 \equiv (\theta_p - \theta_0)^T \mathcal{C}_p^{-1} (\theta_p - \theta_0)
\end{align}
This quantifies the distance, within the Gaussian approximation, of the distribution with respect to a given point and is a natural quantity to consider since the PC modes are maximizing it, as in \cref{Eq:PCAOptimization}.

This quantity is linearly invariant, under the transformation in \cref{Eq:LinearReparametrization}:
\begin{align}
{\rm SNR}^2 =& (\tilde{\theta}_p -\tilde{\theta}_0)^T A^{-T} A^T \tilde{\mathcal{C}}_p^{-1} A A^{-1} (\tilde{\theta}_p -\tilde{\theta}_0) \nonumber \\
=& (\tilde{\theta}_p -\tilde{\theta}_0)^T \tilde{\mathcal{C}}_p^{-1} (\tilde{\theta}_p -\tilde{\theta}_0)
\end{align}
In the previous expression we have made explicit the reference point to highlight that SNR is also invariant under the broader family of affine transformations, including parameter translations, once the reference is specified. From here onward we fix $\theta_0=0$ for conciseness.

We can now write the decomposition of SNR in PC modes, given that they are orthogonal in the given parameter basis:
\begin{align}
{\rm SNR}^2 = \sum_j \lambda_j (Q_j^T \theta_p)^2 = \sum_j \lambda_j q_j^2
\end{align}
where $q_j \equiv Q_j^T \theta_p$ is the projection of the parameter vector on the $j$-th principal component and $\lambda_j$ is the corresponding eigenvalue of the Fisher matrix.
While this looks promising since the total SNR is parameter invariant, we now show that its decomposition in PC modes is not.
Using \cref{Eq:PCA} and its transformed analog:
\begin{align} \label{Eq:PCA_transformed}
\tilde{\mathcal{F}}_p \tilde{Q} = \tilde{Q} \tilde{\Lambda}
\end{align}
along with \cref{Eq:Cp_tilde}, it can be shown that:
\begin{align} \label{Eq:PCA_relation}
(A^{-T}Q) \Lambda (A^{-T}Q)^T = \tilde{Q} \tilde{\Lambda} \tilde{Q}^T
\end{align}
It follows that each SNR component in the transformed basis is equal to:
\begin{align}
\tilde{\rm SNR}_j^2 \equiv & \tilde{\lambda}_j (\tilde{Q}_j^T \tilde{\theta}_p)^2 = (\tilde{Q}_j A^{-T} Q \Lambda Q^T \theta_p) (\tilde{Q}_j^T A \theta_p) \nonumber \\
= & \sum_k \lambda_k  (Q_k^T\theta_p^T)(Q_k^T A^{-1} \tilde{Q_j})(\tilde{Q_j}^T A \theta_p) \nonumber \\
\neq & {\rm SNR}_j^2 \equiv \lambda_j (Q_j^T \theta_p)^2
\end{align}
This expression involves a sum over all the eigenvalues in the original basis and, since $A$ is not necessarily an orthogonal transformation, it cannot be reduced to the j-th component of the SNR in the original basis.
This means that the identification of most discrepant modes depends on the parameter basis that is used.

We note here that, by using \cref{Eq:Cp_Def} and \cref{Eq:Thetap_Def}, we can rewrite \cref{Eq:SNR} as:
\begin{align} \label{Eq:SNR_Dec2}
{\rm SNR}^2 = & \theta_d^T \mathcal{C}_d^{-1} \theta_d + \theta_{\Pi}^T \mathcal{C}_{\Pi}^{-1} \theta_{\Pi} \nonumber \\
+& (\theta_{\Pi}^T - \theta_d^T)(\mathcal{C}_d + \mathcal{C}_{\Pi})^{-1} (\theta_d - \theta_{\Pi}) \,,
\end{align}
where we can see that the posterior SNR contains three contributions: one coming genuinely from the data and other two containing prior terms. 
The first of these two terms gives the SNR of the prior and the second measures the difference between the likelihood and the prior.
While the PCA decomposition provides a way decomposing the posterior SNR as a sum of independent components, it can be verified that the same decomposition would not allow us to isolate its data component.

\section{CPCA decomposition of SNR} \label{Sec:KL.SNR}
Since the CPC modes decorrelate both the posterior and prior they can be used to write a series of useful decompositions.
For starter the $i$-th CPCA projection of the posterior can be written easily in terms of the $i$-th CPCA projection of the prior and maximum likelihood parameters. Defining the projections of KL modes on the parameters as $K_{X,i} = \Psi_i^{-1} \theta_X$ where $X = \{p, d, \Pi\}$:
\begin{align}
K_{p,i} = (1-\lambda^{-1}_{i})K_{d,i} + \lambda^{-1}_{i}K_{\Pi,i} \,,
\end{align}
it follows that the CPCA decomposition of SNR contributions is unique and parameter invariant:
\begin{align}
{\rm SNR}^2 = & (\theta_p^T \Psi^{-T}) \Lambda (\Psi^{-1} \theta_p) = \sum_j \lambda_j (\Psi^{-1}_j \theta_p)^2 \nonumber \\
= & \sum_j \lambda_j [K_p]_{j}^2
\end{align}
each of the terms in the sum is affine invariant since:
\begin{align}
\Psi^{-1}_j \theta_p = \Psi^{-1}_j A^{-1} A \theta_p = \tilde{\Psi}^{-1}_j \tilde{\theta}_p
\end{align}
given that CPC modes are covariant under a reparametrization.

The different contributions to SNR from the data and prior shown in \cref{Eq:SNR_Dec2} can also be written in terms of the CPC modes.
\begin{align} \label{Eq:SNR_Dec_KL1}
{\rm SNR}^2 = & K_d^T (\Lambda - I) K_d + K_{\Pi}^T K_{\Pi} \nonumber\\ 
& + (K_{\Pi}^T - K_d^T)(\Lambda - I)\Lambda^{-1} (K_d - K_{\Pi}) 
\end{align}
and as a sum over independent invariant CPC modes:
\begin{align} \label{Eq:SNR_Dec_KL2}
{\rm SNR}^2 =& \sum_j [2 K_{\Pi,j}^2 - 2 K_{\Pi,j}^T K_{d,j} + \lambda_j K_{d,j}^2 \nonumber \\
&- \frac{1}{\lambda_j} (K_{d,j} - K_{\Pi,j})^2]
\end{align}

Using statistical independence of the CPCs we can isolate the data component of the SNR, as defined by the first term in \cref{Eq:SNR_Dec2}, and write it in terms of quantities that can be easily measured from samples:
\begin{align} \label{Eq:SNR_data_KL}
{\rm SNR}_{d}^2 \equiv & \theta_d^T \mathcal{C}^{-1}_d \theta_d \nonumber \\
= & (K_p^T\Lambda - K_{\Pi}^T )(\Lambda - I)^{-1} (\Lambda K_p -  K_{\Pi}) \nonumber \\
= & \sum_j \frac{1}{{\lambda_j} - 1} (\lambda_j K_{p,j} - K_{\Pi,j})^2
\end{align}
Unlike PCA, in this case, we can isolate the data component of signal-to-noise ratio associated with each CPC mode and that quantity is a parameter invariant.

We conclude this section by pointing out an interesting property of the posterior to prior CPC decomposition.
We consider the Kullback-Leibler divergence of the posterior and the prior:
\begin{align}
D_{KL}(\mathcal{P}||\Pi) = \int \mathcal{P}(\theta) \log\left( \frac{\mathcal{P}(\theta)}{\Pi(\theta)} \right)\, d\theta
\end{align}
which is the posterior average of the data likelihood and quantifies the information gain in going from the prior to the posterior.
In the linear model when all distributions are Gaussian this becomes:
\begin{align}
D_{KL}(\mathcal{P}||\Pi) =& (\theta_\Pi-\theta_p)^T\mathcal{C}_\Pi^{-1}(\theta_\Pi-\theta_p) \nonumber \\
& -\log\det(\mathcal{C}_p\mathcal{C}_\Pi^{-1}) -N +{\rm Tr}(\mathcal{C}_p\mathcal{C}_\Pi^{-1}) \,.
\end{align}
Under data draws from the model evidence the posterior parameters are Gaussian distributed with mean $\theta_\Pi$~\cite{Raveri:2018wln}.
This allows us to compute the average of the Kullback-Leibler divergence under data realizations:
\begin{align}
\langle D \rangle =& 2{\rm Tr}(\mathcal{C}_p\mathcal{C}_\Pi^{-1}) -\log\det(\mathcal{C}_p\mathcal{C}_\Pi^{-1}) -N  \nonumber\\
=& \sum_{j=1}^N \left( 2\lambda_j^{-1} +\log\lambda_j -1 \right)
\end{align}
as a sum over independent CPC modes.

\bibliographystyle{apsrev4-1}
\bibliography{biblio.bib}

\end{document}